\documentclass{aa}
\usepackage{graphicx}
\usepackage{txfonts}
\usepackage{natbib}
\bibpunct{(}{)}{;}{a}{}{,}
\usepackage{color}

\newcommand{\unit}[1]{\ensuremath{\:\mathrm{#1}}}

\newcommand{\modif}[1]{#1}
\newcommand{\modifb}[1]{#1}
\newcommand{\modifbi}[1]{#1}

\begin{document}

\title{On deriving p-mode parameters for inclined solar-like stars}

\author{J. Ballot\inst{1}
\and  T. Appourchaux\inst{2}
\and  T. Toutain\inst{3}
\and  M. Guittet\inst{1}}


\institute{
Max-Planck-Institut f\"ur Astrophysik, Karl-Schwarzschild-Str. 1,
85748 Garching, Germany\\
\email{jballot@mpa-garching.mpg.de}
\and
Institut d'Astrophysique Spatiale, UMR 8617, Universit\'e Paris-Sud, B{\^a}timent 121, 91405 Orsay CEDEX, France\\
\email{thierry.appourchaux@ias.u-psud.fr}
\and
School of Physics and Astronomy, University of Birmingham, Edgbaston,
Birmingham B15 2TT, UK\\
\email{toutain@bison.ph.bham.ac.uk}}

\date{Received 31 December 2007 / Accepted 6 March 2008}

\abstract%
{Thanks to their high quality, new and upcoming asteroseismic observations
  -- with CoRoT, Kepler\modifbi{, and from the ground}... -- can benefit from the experience gained with
  helioseismology.}%
{We focus in this paper on solar-like oscillations, for which the inclination 
of the rotation axis is unknown. 
We present a theoretical study of the errors of p-mode parameters determined 
by \modifb{means} of a maximum-likelihood estimator, and we also analyze correlations 
and biases.}%
{\modifb{W}e have used different, complementary approaches: we have performed
either semi-analytical computation of the Hessian matrix, fitting of
single mean profiles, or Monte Carlo simulations.}%
{We give first analytical approximations for the errors of frequency, inclination and rotational splitting. The determination of the inclination is very challenging for the common case of slow rotators (like the Sun), making difficult the determination of a reliable rotational splitting. Moreover, due to the numerous correlations, biases -- more or less \modif{significant} -- can appear in the determination of various parameters in \modifb{the} case of bad inclination fittings, especially when a locking at 90\degr\ occurs. This issue concerning inclination locking is also discussed. Nevertheless, the central frequency and some derived parameters \modifb{such as} the total power of the mode are \modifb{free} of such biases.}%
{}

\maketitle

\section{Introduction}\label{Sec:Intro}
A new era of the asteroseismic observations is now opening with CoRoT \citep{BaglinM06_COROT}, and will be pursued in the near future with Kepler \citep{BoruckiK07_KEPLER}. As these missions will provide long and uninterrupted time series of intensity measurements, the quality of these data will become closer to those of helioseismology, although with a higher level of noise \modifb{expected}. In such a context, asteroseismology can fruitfully inherit the methods and techniques developed for helioseismology, especially for so-called `Sun-as-a-star' observations, \modifb{such as} those provided by the \modifb{space} instruments GOLF \citep{GabrielG95_GOLF} or VIRGO \citep{FrohlichR95_VIRGO} \modifb{onboard} the SoHO spacecraft or by ground-based networks like BiSON \citep{ChaplinE96_BISON}.

We discuss in this paper the usual way to extract information on stochastically excited acoustic (p) modes \modifb{in helioseismology}. P modes can be described by several parameters (frequency, lifetime, amplitude... ) which can be derived by fitting the stellar oscillation power spectrum with a maximum-likelihood estimation. As for any \modifb{measurement}, we need to associate \modifb{with} each derived parameter a correct error bar, \modifb{that} is essential to estimate the \modifb{significance} of the measurement and thus to be able to do a reasonable interpretation.
We present here theoretical results on the derivation of errors of p-mode parameters and their correlations. This work generalizes some results already known in helioseismology \citep{Libbrecht92,ToutainA94} by adding the inclination of the stellar rotation axis, $i$, as an extra free parameter.

Fitting methods with a free inclination angle \modifb{have} first been analyzed by \citet{GizonS03}, then by \citet{BallotG06}. In the present paper, we develop a more complete version of a preliminary work \citep{BallotAT07} on analytical formulations of error bars for several parameters, and we have completed it with an analysis of correlations and biases. We have especially considered here the very common cases of slow rotators, for which the mode linewidth is greater than (or similar to) the rotational splitting, \modifb{giving rise to} a blending of multiplet components.

\modifb{In \S~\ref{Sec:Methods}, w}e define the model assumed for p modes and the fitting method. Section~\ref{Sec:Error} deals with error bars obtained for the mode frequencies, the splittings and the inclination. The correlation between different parameters \modifb{is} studied in \S~\ref{Sec:Correl} and some biases of the method are analyzed in \S~\ref{Sec:Biases}. Finally we briefly discuss the locking of the inclination determination at 90\degr\ which often appears during fitting (\S~\ref{Sec:Lock}).

\section{Models and Methods}\label{Sec:Methods}

\subsection{Modeling the power spectrum}\label{SSec:Model}
\modifb{S}tellar acoustic eigenmodes are characterized by their degree $l$, their azimuthal order $m$ and their radial order $n$.
In this study, we treat p modes according to the solar paradigm. Modes are modelled as stochastically excited and intrinsically damped harmonic oscillators \citep[see][]{KumarF88}. In that case, the power spectrum -- obtained by computing the discrete Fourier transform of an evenly sampled timeseries -- of such modes is distributed around a mean Lorentzian profile with an exponential probability distribution.
A Lorentzian profile is defined as
\begin{equation}
L(A,\nu_0,\Gamma; \nu)=\frac{A}{1+\left(\frac{\nu-\nu_0}{\Gamma/2}\right)^2},
\end{equation}
where $A$ is the mode height, $\Gamma=(\pi\tau)^{-1}$ the mode linewidth, linked to the damping time $\tau$, and $\nu_0$ the mode frequency.
For our examples we have considered the value $\Gamma=1\unit{\mu Hz}$ in the whole study, that corresponds to a lifetime around 3--4 days. This is a typical observed value of solar p-mode linewidths in a broad range around 2500--3000\unit{\mu Hz} \citep[e.g.][]{GarciaJR04SoHO}.

Since stellar rotation lifts the azimuthal degeneracy of eigenmodes, 
the power spectrum of a mode with a degree $l$  is a multiplet with $2l+1$ components. Extrapolating from the solar case, we assume equipartition of energy between the components in a multiplet, and we define a multiplet as a symmetric profile:
\begin{equation}
M_l (A, \nu_0, \Gamma, \nu_s, i; \nu) =
\sum_{m=-l}^l a_{l, m}(i) L(A,\nu_0+m\nu_s, \Gamma ; \nu),
\end{equation}
where $\nu_s$ is  the rotational splitting and the inclination $i$ is the angle between the rotation axis and the line of sight.

We have used here the approximation of \citet{Ledoux51} for an uniform rotation: the frequency shift of the $m$-component relative to the central one is $\nu_{m}-\nu_0=m\nu_s$; moreover, we consider $\nu_s\approx\Omega/(2\pi)$, that is the asymptotic regime at high order $n$ ($\Omega$ is the stellar rotation rate). This approximation is valid \modifb{when} rotation is sufficiently slow and there is \modifb{neither a} strong differential rotation \modifb{nor a} strong magnetic activity.

Next, the inclination acts only on $a_{l,m}(i)$, the amplitude ratios of components inside a multiplet, which \modifb{satisfies} the relationship $\sum_m a_{l, m} = 1$. Thus the total power of a multiplet is always $P=\frac{\pi}{2}A\Gamma$. The amplitude ratios are written
\begin{equation}
a_{l, m}(i)=\frac{|l-m|!}{|l+m|!}(P_{l, m}(\cos i))^2
\end{equation}
where $P_{l, m}$ are the associated Legendre functions \citep[see][]{GizonS03}. To derive this expression, when the flux is integrated over the full stellar disc, we need to assume that the weight\modifb{ing} function, which gives the contribution of a point on the disc to the integral, is a function of the distance to the disc center only. This is correct for intensity measurements, because the weight\modifb{ing} function is mainly linked to the limb-darkening. However for velocity measurements, we can observe departures from this law, since the rotation of the star induces an asymmetry in the velocity field.

Last, we have also assumed that a mode is not correlated with any other mode\modifb{s} or with the convective background noise. Doing so we neglect in the present study any possible asymmetry of the Lorentzian profiles \citep{NigamKSS98}. \modif{According to what we have learned from helioseismology, neglecting asymmetries could introduce systematic errors (i.e. biases) in mode frequency determination. These errors are
of the order of 0.1\unit{\mu Hz} in the solar case \citep{ToutainA97}. This is
on par with the statistical error (i.e. the standard deviation) of the frequency for time series of
several months. For longer time series (few years), the asymmetry should be included in the fitted profiles to avoid systematic errors in mode frequencies.}

To summarize, we have considered that the star is observed in intensity, that the mode excitation mechanisms are close to those of the Sun, and that the star does not rotate too \modifb{rapidly} (\modifb{a} few times the solar rate).

\subsection{Maximum-likelihood estimator}\label{SSec:MLE}
P-mode parameters are fitted with a classical maximum-likelihood estimation technique and the associated error and correlation are estimated by inverting the Hessian matrix \citep{ToutainA94,AppourchauxG98}.
In practice, instead of maximizing the likelihood, we minimize the negative logarithm of the likelihood function, which \modifb{yields} for a random exponential noise:
\begin{equation}
\ell(\vec{\lambda})=-\ln {\cal L}(\vec{\lambda})=\sum_{k=1}^N \ln S(\vec{\lambda}; \nu_k) + \frac{S_k}{S(\vec{\lambda}; \nu_k)}
\end{equation}
where $S$ is the model of the spectrum measured $\{S_k\}_{k=1,N}$ at frequencies $\{\nu_k\}$, and
$\vec{\lambda}=(\lambda_1,\dots,\lambda_p)$ is the set of $p$ parameters to \modifb{be adjusted}.
We denote hereafter $\tilde\lambda_j$ an estimation of $\lambda_j$.

As the observed intensity fluctuations are integrated over the whole stellar disc, only low-degree modes are visible. A quick estimation of their visibility indicates that we will mainly be able to detect only modes with $l \le 2$ and perhaps a few $l=3$ modes.

We have therefore considered the following fitting cases:
1) we fit a mode $l=1$ alone; $S$ is described by 6 parameters: $A$, $\nu_0$, $\Gamma$, $i$, $\nu_s$, and a additive background $B$ assumed to be flat \modifb{within} the fitting window. \modifb{In practice}, we fit the \modifb{logarithm} of some \modifb{of the} parameters: $a=\ln A$, $\gamma=\ln\Gamma$ and $b=\ln B$. 2) We fit a pair of modes $l=0$ and 2; assuming a common linewidth for both modes $S$ is described by 8 parameters; 3) We fit a sequence of modes $l=0$, 2 and 1; $S$ is described by 11 parameters, assuming the splitting is the same for the consecutive $l=1$ and 2 modes.

The \modif{standard deviations (error bars)} $\sigma_j$ associated \modifb{with} $\lambda_j$ are estimated with the covariance matrix $\mathbf{C}$, computed by inverting the Hessian matrix $\mathbf{H}$ ($\mathbf{C}=\mathbf{H}^{-1}$). The diagonal elements of  $\mathbf{C}$ give the errors $c_{jj}=\sigma_j^2$, while the \modifb{non}-diagonal elements give the covariances $c_{ij}=\sigma_{ij}=
\rho_{ij}\sigma_i\sigma_j$ ($\rho_{ij}$ are the correlation coefficients).

The terms of the Hessian are:
\begin{equation}
h_{ij}=\left.\frac{\partial^2 \ell(\vec{\lambda})}{\partial \lambda_i \partial \lambda_j}\right|_{\vec{\lambda}=\vec{\tilde\lambda}}.
\end{equation}

Following \citet{Libbrecht92} and \citet{ToutainA94}, we define a theoretical Hessian corresponding to the average of a large number of realizations:
\begin{equation}\label{eq:hesssum}
h_{ij}=\sum_{k=1}^{N}\frac{1}{S^2(\vec{\lambda};\nu_k)} \frac{\partial S}{\partial \lambda_i}\frac{\partial S}{\partial \lambda_j}.
\end{equation}
If the frequency bin is much smaller than the mode linewidth, we can approximate \modifb{the Hessian} by the integral:
\begin{equation}\label{eq:hessint}
\ h_{ij}=T \int_{-\infty}^{+\infty}\frac{1}{S^2(\vec{\lambda};\nu)} \frac{\partial S}{\partial \lambda_i}\frac{\partial S}{\partial \lambda_j} d\nu,
\end{equation}
where $T$ is the observation duration.

In the next sections, results concerning errors, correlations and biases are obtained with semi-analytical computations of Eqs.~\ref{eq:hesssum} and \ref{eq:hessint}, by fitting the mode profile model as in \citet{ToutainE05}, or with Monte Carlo simulations.

\section{Theoretical error bars}\label{Sec:Error}

\subsection{Error of the central frequency $\nu_0$}\label{SSec:Errnu0}

\begin{figure*}[!ht]
\centering
\includegraphics[width=8cm]{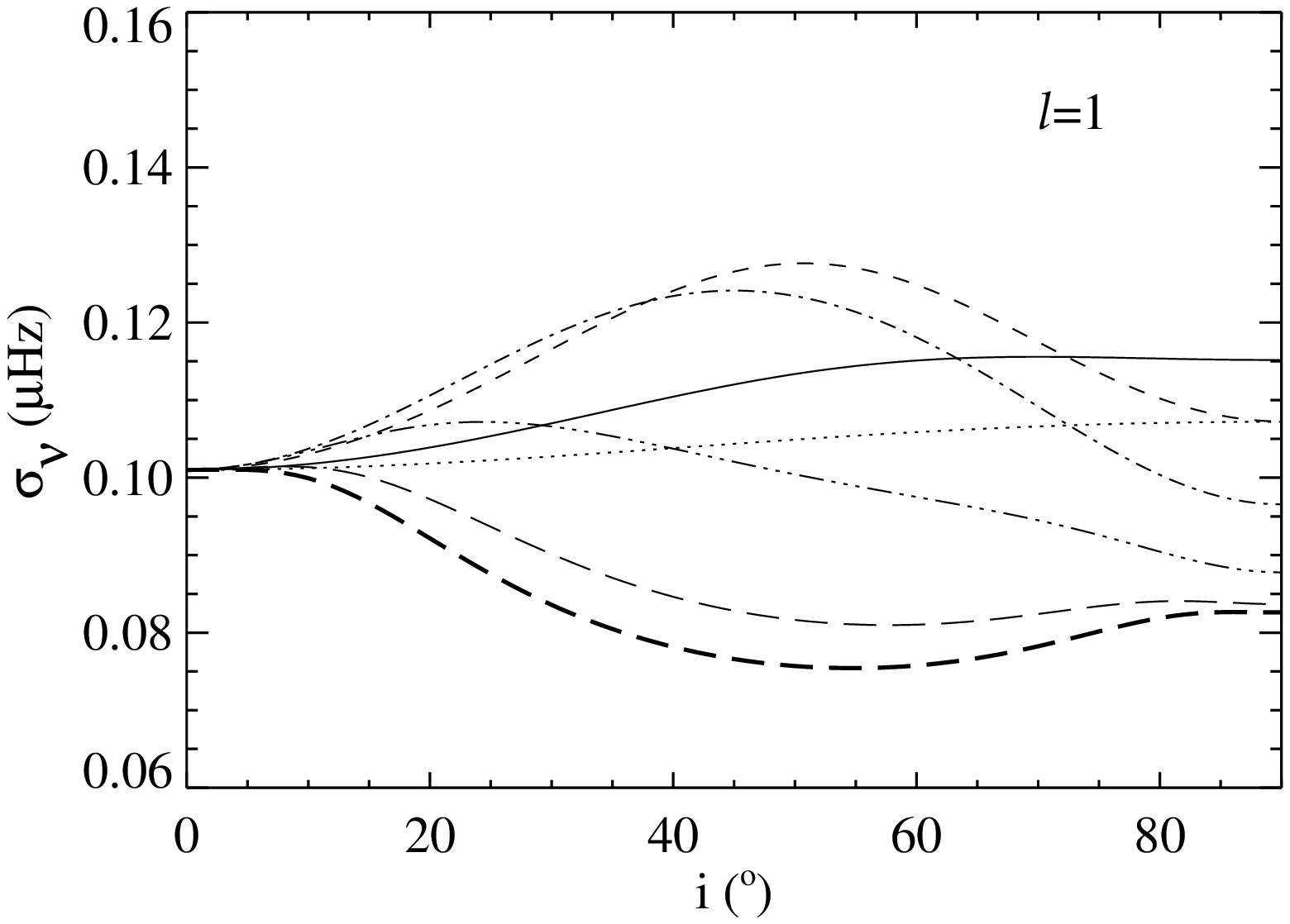}\includegraphics[width=8cm]{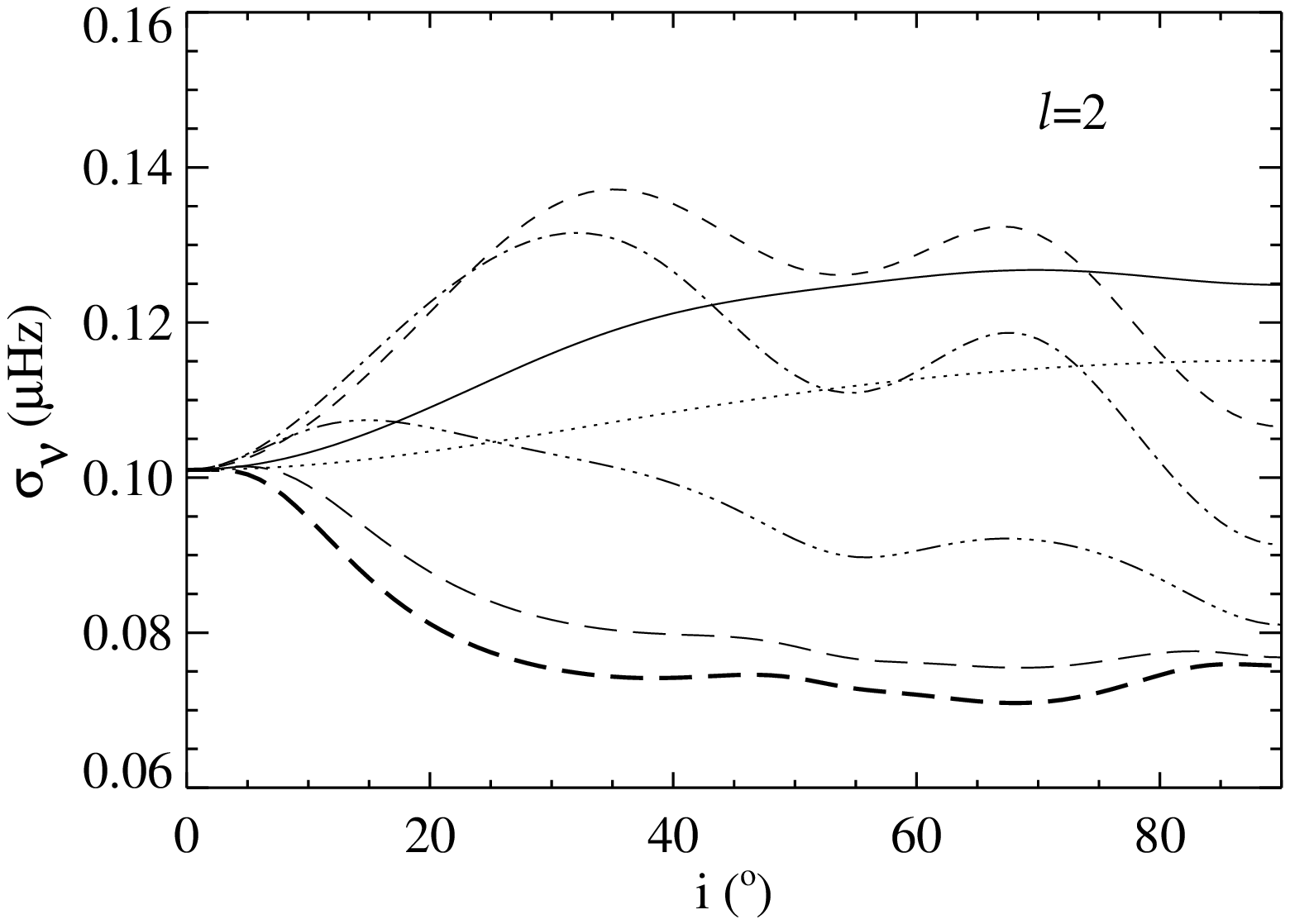}
\caption{
Frequency error computed with Eq.~(\ref{eq:errnu}) for modes $l=1$ (\textit{left}) and $l=2$ (\textit{right}) \modifb{as a function of} the angle $i$, for different reduced splittings $x_s=$ 0.4 (dots), 0.8 (solid line), \modif{1.6 (dashes), 2.4 (dot-dash), 4.0 (dot-dot-dot-dash), 8.0 (long dashes); that corresponds, for $\Gamma=1\:\mu$Hz, to $\Omega=$0.5, 1, 2, 3, 5, 10$\:\Omega_\odot$}. Thick long dashes indicate the limit when $x_s \gg 1$, computed with Eq.~(\ref{eq:errnu_as}). Here $\beta=1/20$ and $T=6\:$months.}\label{Fig:errnu}
\end{figure*}

We generalize in this section the results of \citet{Libbrecht92} and \citet{ToutainA94}, inferred for $l=0$ and for multiplets in the solar configuration ($i=90\degr$).

Considering an isolated multiplet $S=B+M_l$, we can easily verify that $h_{\nu_0,j}(=h_{j,\nu_0})=0$ except for $\lambda_j=\nu_0$. Since $S$ is even \modifb{in} the variable $\nu-\nu_0$, the derivatives $\partial_{\nu_0}S=-\partial_{\nu}S$ are odd. However we can verify that $\partial_{\lambda_j}S$ is even relative to the variable $\nu-\nu_0$ for all of the other parameters $\lambda_j=B,A,\Gamma,\nu_s$ and $i$. This is easy to demonstrate by noting that any modification of one or several of these parameters preserves the symmetry of the profile \modif{with} respect to $\nu-\nu_0$. With the same argument, we easily demonstrate that, after any variable change $\lambda_j=f(B,A,\Gamma,\nu_s,i)$, $\partial_{\lambda_j}S$ is still even.
Thus, in Eq.~\ref{eq:hessint}, the function under the integral is odd, and the integral vanishes.
We deduce that
\begin{equation}
\sigma^2_{\nu_0}=h_{\nu_0\nu_0}^{-1}
\qquad\mbox{and}\qquad
\sigma_{\nu_0j}=0 \quad \forall\lambda_j\neq\nu_0.
\end{equation}
The \modifb{lack of} correlation of $\nu_0$ with all other parameter is \modifb{well established} in Monte Carlo (MC) simulations. When a pair (for instance $l=0$ and 2) is fitted, this is still correct while \modifb{the} modes are well separated: given $\lambda_j^{(0)}$ and $\lambda_k^{(2)}$ two parameters of the considered $l=0$ and $l=2$ modes, then $\partial_{\lambda_j^{(0)}}S \partial_{\lambda_k^{(2)}}S \approx 0\ \forall\nu$, since either $\partial_{\lambda_j^{(0)}}S$ or $\partial_{\lambda_k^{(2)}}S$ vanishes, as modes are sufficiently \modifb{far} apart. However, when modes are blended, \modifb{crosstalk} can occur and the central frequency can become correlated with other parameters, especially with the central frequency of the neighbor mode. Moreover, if the mode profiles are not symmetric (asymmetries in the Lorentzian or in the splittings), the mode frequencies can also become correlated with the others parameters, but \modifb{\textit{a priori} we expect negligibly small} effects.

From Eq.~\ref{eq:hessint}, we derive the error for $\nu_0$:
\begin{equation}
\sigma^2_{\nu_0}=\frac{1}{4\pi}\frac{\Gamma}{T}f_l(\beta,x_s,i),\label{eq:errnu}
\end{equation}
with $\beta=B/A$ the noise-to-signal ratio and $x_s=2\nu_s/\Gamma$ the \textit{reduced splitting}.
For $l=0$, we find the formula of \citet{Libbrecht92}:
\begin{equation}
f_0(\beta\mbox{ only})=\sqrt{\beta+1}\left(\sqrt{\beta+1}+\sqrt{\beta}\right)^3.
\end{equation}
For $l>0$, the simplest form of $f_l$ is its integral form:
\begin{equation}
f_l(\beta,x_s,i) =
\frac{\pi}{4}
\left[
\int_0^{+\infty} \left[
\frac{\sum_{m} a_{l,m}(i) (x+mx_s) L_r^2(x+mx_s)}%
{\beta+\sum_{m} a_{l,m}(i)L_r(x+mx_s)}
\right]^2 dx
\right]^{-1}
\end{equation}
where
\begin{equation}
L_r(x)=\frac{1}{1+x^2}
\end{equation}
is the reduced Lorentzian.

Figure~\ref{Fig:errnu} shows the evolution of $\sigma_{\nu_0}$ with $i$ and $x_s$ for $l=$1 and 2 modes.
As expected, $f_l(\beta,x_s,i)$ \modifb{approches} $f_0(\beta)$, as $i$ or $x_s$ \modifb{approches} zero.
We can also notice that, when $x_s \ll 1$, the splitting acts as an extra width, increasing the error for $\nu_0$. Furthermore we clearly see that, depending of the values of $i$ and $x_s$, the error bars for $\nu_0$ can vary within a factor of 2.
When $x_s \gg 1$, i.e. for large rotation rates, the components are well-separated, the fitting gives \modifb{exactly} the same result as when each $m$-component is independently fitted and the central frequency is computed with a weighted average: $\nu_{0}=\sum_{m}{\nu_m}{\sigma_{\nu_m}^{-2}}/\sum_{m}{\sigma_{\nu_m}^{-2}}$. The associated error is then
\begin{equation}
\sigma^{-2}_{\nu_0}=\sum_{m=-l}^l\sigma^{-2}_{\nu_m}=\frac{4\pi T}{\Gamma}\sum_{m=-l}^l\left[f_0\left(\frac{\beta}{a_{l, m}(i)}\right)\right]^{-1}.
\label{eq:errnu_as}
\end{equation}
This value, plotted in Fig.~\ref{Fig:errnu} as a thick dashed line, gives the lowest limit for the error at \modifb{a} given angle, linewidth and S/N ratio.

\subsection{Error of the angle $i$ and the splitting $\nu_s$}\label{SSec:Errnui}
\begin{figure*}[!ht]
\centering
\includegraphics[width=7.5cm]{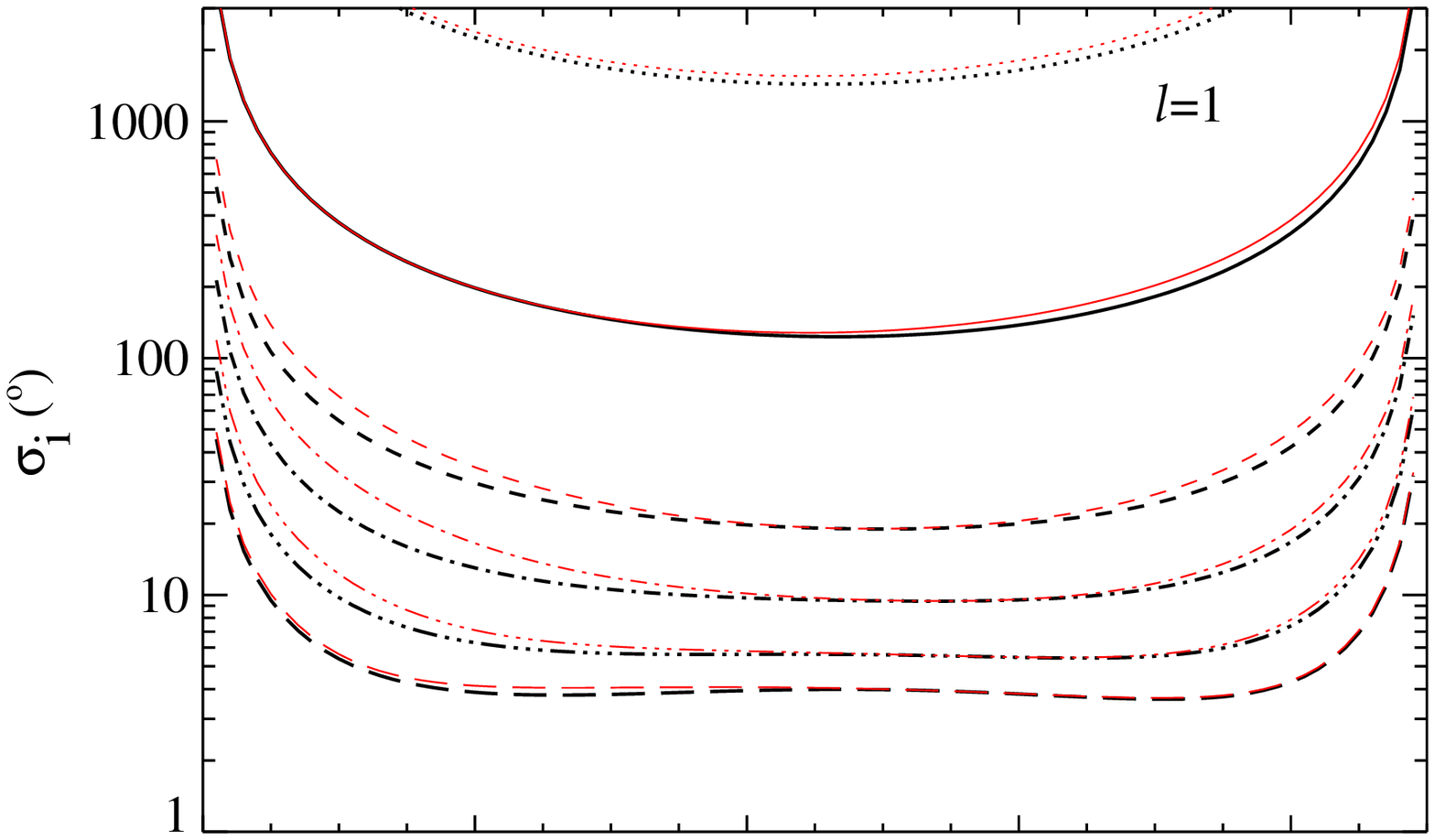}\hspace*{-1cm}\includegraphics[width=7.5cm]{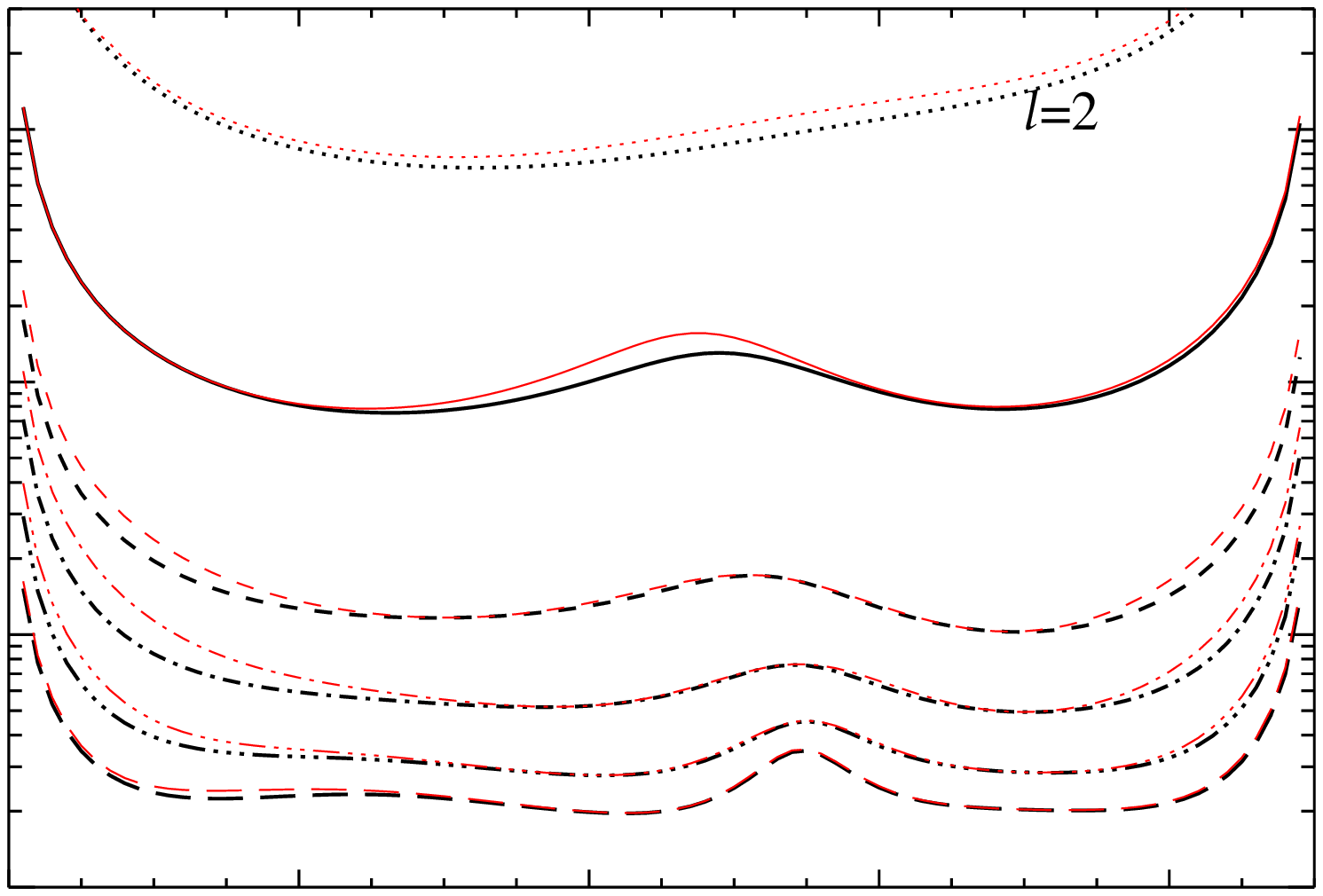}\\
\vspace*{-1cm}\includegraphics[width=7.5cm]{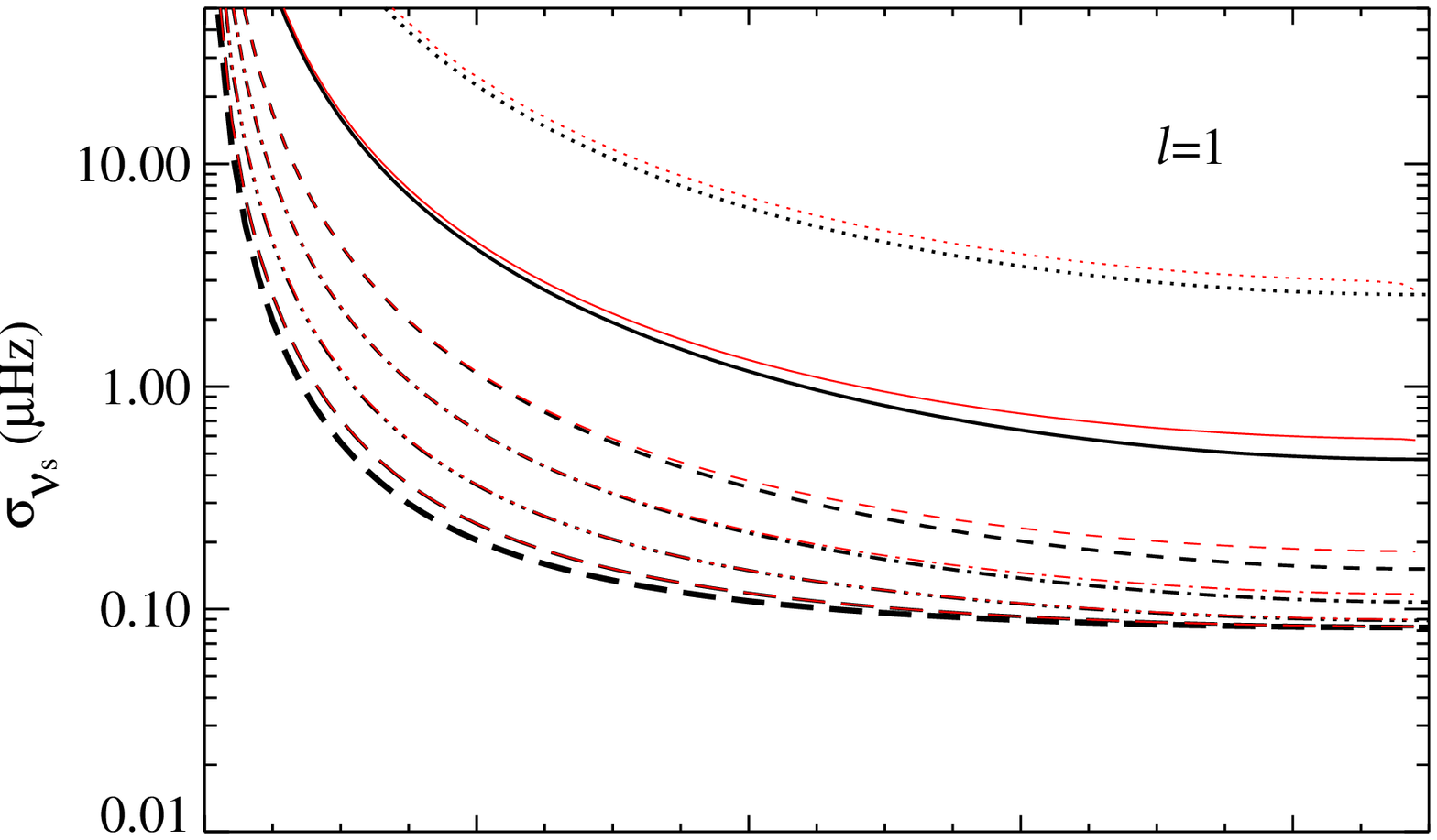}\hspace*{-1cm}\includegraphics[width=7.5cm]{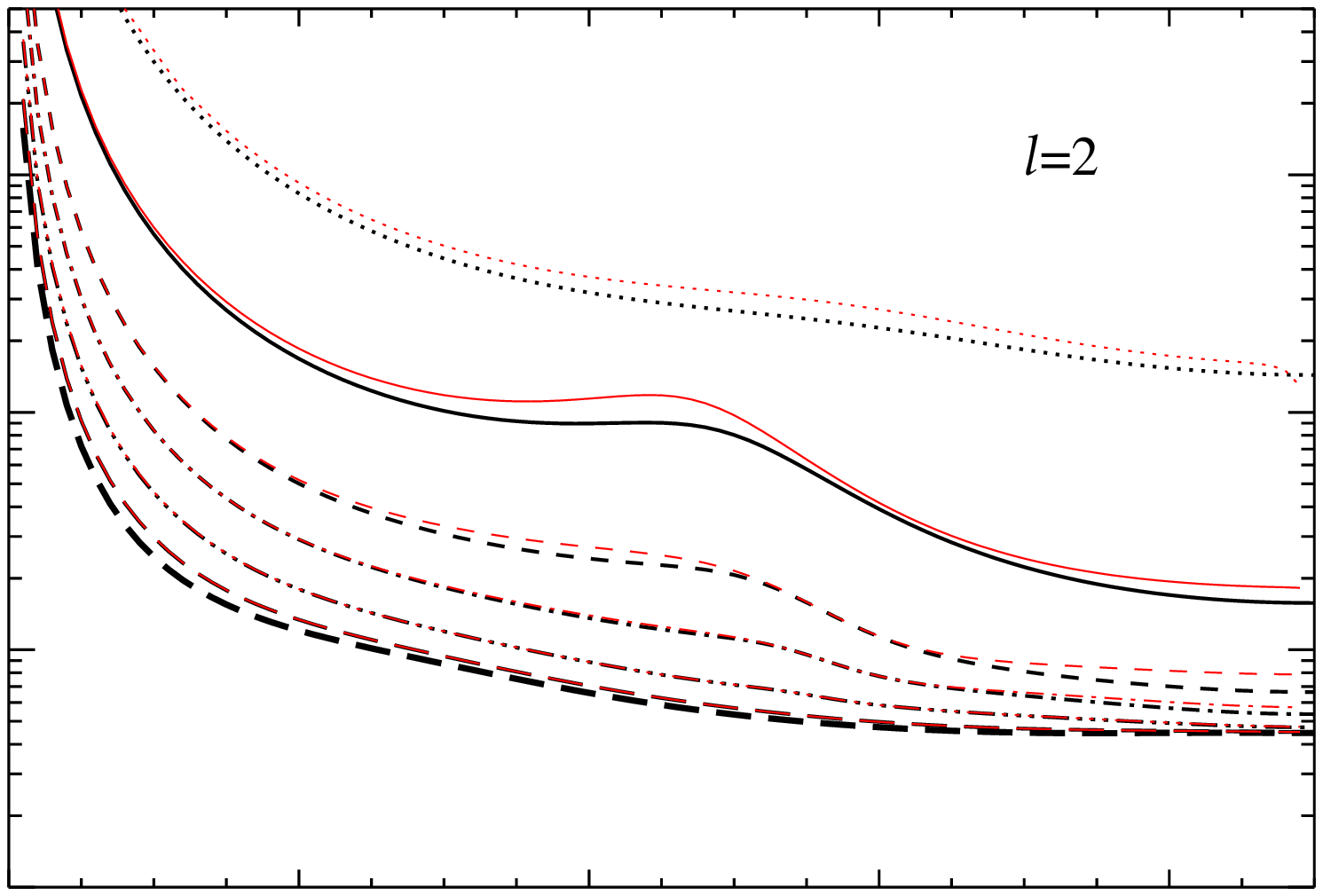}\\
\vspace*{-1cm}\includegraphics[width=7.5cm]{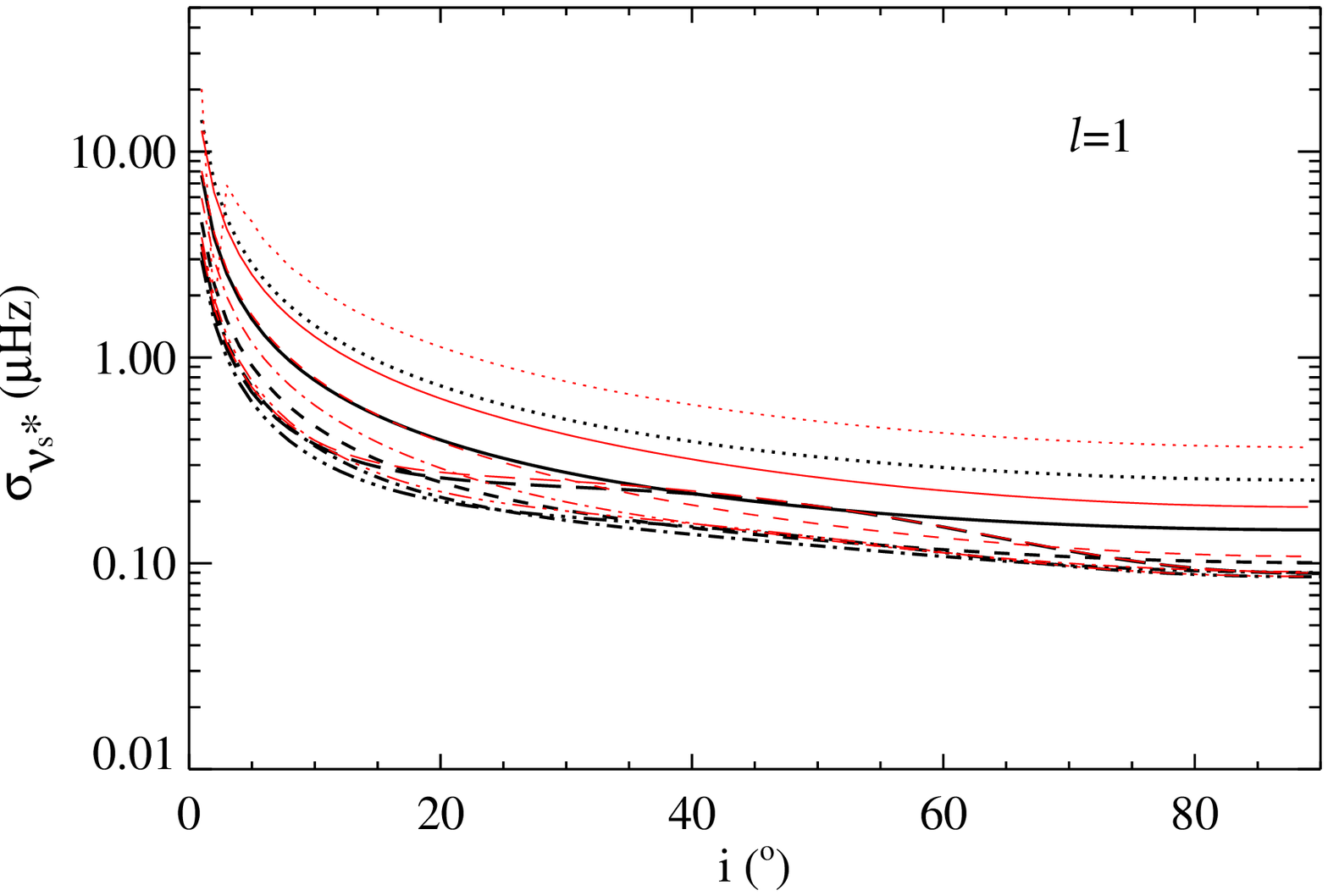}\hspace*{-1cm}\includegraphics[width=7.5cm]{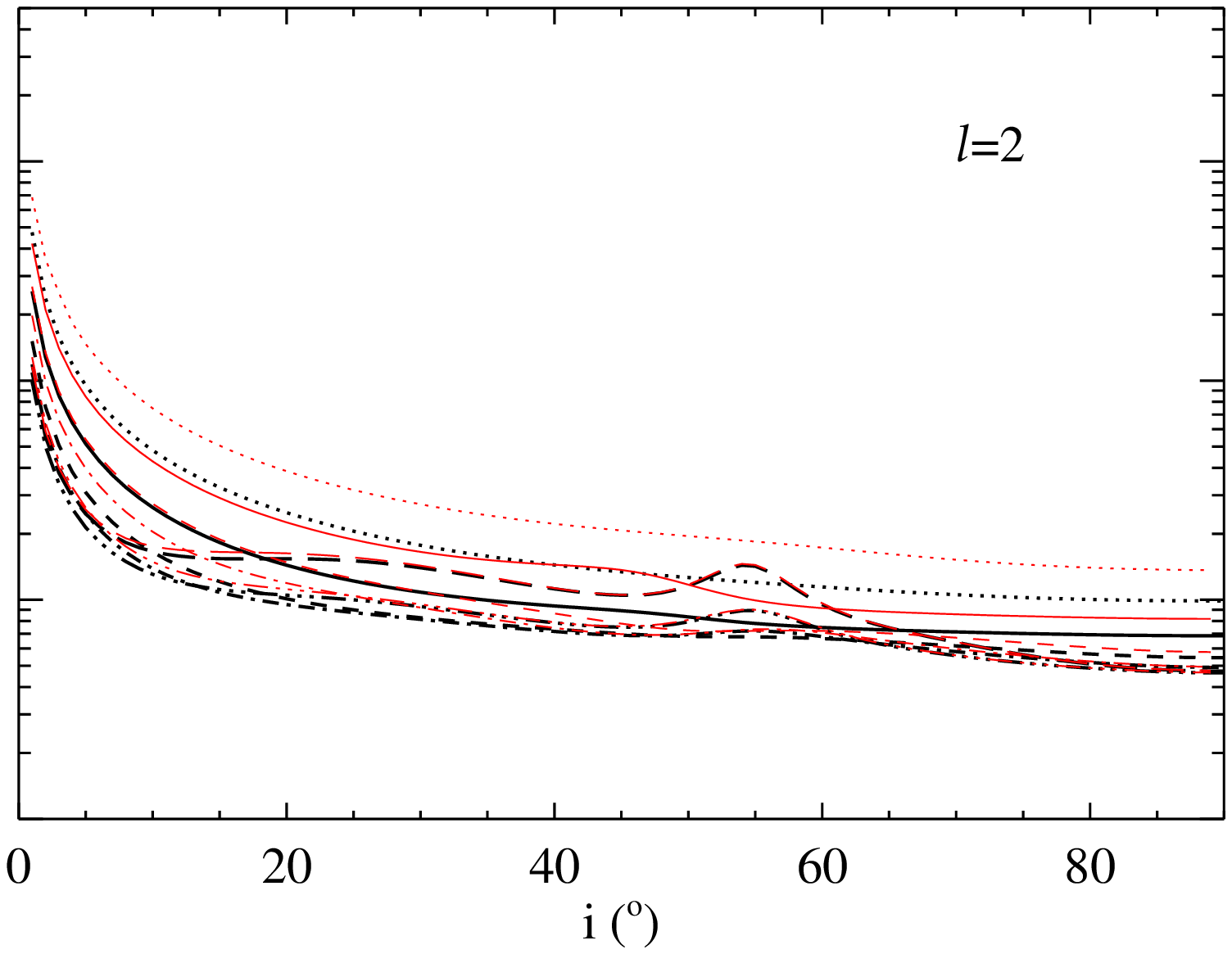}
\caption{Error for the angle $i$, the splitting $\nu_s$, and the projected splitting $\nu_s^*$, as a function of $i$, for modes $l=1$ (\textit{left}) and $l=2$ (\textit{right}), for different values of $x_s$ (see caption of Fig.~\ref{Fig:errnu}). Black lines show errors obtained with the simplified Hessian, thin red lines are obtained by inverting the full Hessian. Thick long dashes \modifb{in the} middle panels indicate the limit of the error on $\nu_s$ when $x_s \gg 1$, computed with Eq.~(\ref{eq:errnus_as}). Here $\beta=1/20$ and $T=6\:$months.}\label{Fig:errspl}
\end{figure*}

We focus now on the error of the inclination and the splitting, or other related variables. \cite{BallotG06} have used the \modifb{pair} of parameters $(i,\nu_s^{*})$ instead of $(i,\nu_s)$, where 
$\nu_s^{*}=\nu_s\sin i$ is the \textit{projected splitting}.
The authors did not see any difference in the determination of $i$ using one set of variables or the other. 
We analytically derive this result here, and we study more generally what happens when we fit other sets of parameters, for instance $(\sin^2 i,\nu_s)$ as further mentioned in \S~\ref{Sec:Lock}.

Let us denote \modifb{by} $(p,q)$ a new \modifb{pair} of parameters such $i=f_i(p)$ and $\nu_s=f_S(p,q)$.
We consider a simplified Hessian matrix 2$\times$2 with elements (see Eq.~\ref{eq:hesssum}):
\begin{eqnarray}
h_i&=&\sum \frac{1}{S^2}\left(\frac{\partial S}{\partial i}\right)^2\\
h_{s}&=&\sum \frac{1}{S^2}\left(\frac{\partial S}{\partial \nu_s}\right)^2\\
h_{is}&=&\sum \frac{1}{S^2}\frac{\partial S}{\partial \nu_s}\frac{\partial S}{\partial i}
\end{eqnarray}
The \modif{errors} for $i$ and $\nu_s$ are
\begin{eqnarray}
\sigma_i^{-2}&=&h_i-\frac{h_{is}^2}{h_s},\\
\sigma_{\nu_s}^{-2}&=&h_s-\frac{h_{is}^2}{h_i}.
\end{eqnarray}
We denote \modifb{by} $\mathbf{G}$ the simplified Hessian matrix \modifb{for} the new parameters. Its elements are $g_p$, $g_q$ and $g_{pq}$. Using rules of (partial) derivatives, we show that:
\begin{equation}
\sigma_p^2=\frac{g_q}{g_pg_q-g_{pq}^2}=\left|\frac{dp}{di}\right|^2\frac{h_s}{h_ih_s-h_{is}^2}=\left|\frac{dp}{di}\right|^2\sigma_i^2,
\end{equation}
and
\begin{equation}
\sigma_q=\left(\frac{\partial f_S}{\partial q}\frac{d f_i}{d p}\right)^{-1} \sqrt{\frac{g_p}{h_i}}\sigma_{\nu_s}.
\end{equation}
\modif{Specifically} for $p=i$ and $q=\nu_s^{*}$, we find:
\begin{equation}
\sigma_{\nu_s^*}=\sigma_{\nu_s}\sin i\sqrt{1+\frac{\nu_s^2}{\tan^2 i}\frac{h_s}{h_i}-2\frac{\nu_s}{\tan i}\frac{h_{is}}{h_i}}.
\end{equation}

First we conclude that the error for $i$ does not depend on the choice made for the other variables: $\nu_s$, $\nu_s^{*}$, or another combination. MC simulations and all of our other tests confirm this. Second, using $\cos i$ or $\sin i$ does not modify the error: we retrieve natural relationships such as $\sigma_{\sin^2 i}=|\sin 2i|\sigma_{i}$.

Figure~\ref{Fig:errspl} shows the errors we derive for the angle $i$, the
splitting $\nu_s$ and the projected splitting $\nu_s^{*}$ for 6-month long
observations. We compare the results from \modifb{a} simplified Hessian to those obtained
by fitting directly the profile model \citep[see method in ][]{ToutainE05}.
For the angle, both computations give very similar results. Estimated errors clearly demonstrate the difficulty \modifb{in deriving} a reliable determination of the angle $i$ from one single mode when the rotation is \modifb{less than} 2$\Omega_{\sun}$: the uncertainty covers almost the \modifb{entire} possible range. 
$l=2$ modes provide slightly better estimates of $i$ than $l=1$ modes because they have more components and \modifbi{the displacement of the $l=2$ modes is twice that of the $l=1$ modes}. These error estimations are in agreement with those of \citet{GizonS03}, who have also numerically derived errors from the Hessian.
Dividing by $\sqrt{8}$, the values of the errors in Figure~\ref{Fig:errspl} would give the corresponding errors for 4 years of observation (scales \modif{as the square root of the ratio of the observing times}). Unless several modes are used 
to determine the inclination and the rotation, 4 years is still not sufficient to get enough accuracy when $\Omega\la\Omega_{\sun}$. However, if we measure 10 independent modes (for instance, 5 $l=1$ and 5 $l=2$), by averaging the results, we can expect to reduce the error to $\sim15\degr$ for $\Omega=\Omega_{\sun}$.

Concerning the error for the splitting $\nu_s$, our simplified analysis \modifb{also gives} a result in good agreement with the complete numerical computation. We notice as well the dramatic \modifb{increase} of the errors when $\nu_s$ decreases. As previously done for the central frequency, we can also derive here the limit of \modifb{the} splitting error when $x_s \gg 1$ and that all of the components can be fitted independently.
The resulting error for $\nu_s$ in this case is given by:
\begin{equation}
\sigma^{-2}_{\nu_s}=\sum_{m=-l}^l m^2\sigma^{-2}_{\nu_m}
=\frac{8\pi T}{\Gamma}\sum_{m=1}^l m^2\left[f_0\left(\frac{\beta}{a_{l, m}(i)}\right)\right]^{-1}.
\label{eq:errnus_as}
\end{equation}
Interestingly enough it gives a lower bound for $\sigma_{\nu_s}$ for given $i$, $T$, $\Gamma$ and $\beta$.

\modifb{The lower} plots in Fig.~\ref{Fig:errspl} show the error of the projected splitting $\nu_s^{*}$. One see\modifb{s} that our simplified calculation is slightly \modifb{more crude} in this case. In this simplified approach, the correlations of $i$, $\nu_s$ and $\nu_s^{*}$ with the other parameters are neglected. They are actually non\modifb{-}negligible, especially \modifb{for} the linewidth $\Gamma$. For slow rotation, $\nu_s^{*}$ is a bit more correlated with the linewidth $\Gamma$ than $\nu_s$ (see \S~\ref{Sec:Correl}), that mainly explains why the discrepancy between simplified and full Hessian computations is larger for $\nu_s^{*}$ than for $\nu_s$.
Nevertheless, the most remarkable fact is the smaller \modifb{scatter} of the curves compared to the previous plots: as suggested by \citet{BallotG06}, this computation indeed shows that $\sigma_{\nu_s^{*}}$ is noticeably less sensitive than $\sigma_{\nu_s}$ to the value of $\nu_s$.

\section{Correlations}\label{Sec:Correl}

We discuss in this section the correlations between the different parameters of a mode and \modifb{the effects of the correlations} on data analysis.


\begin{figure*}[!ht]
\centering
\includegraphics[width=15cm]{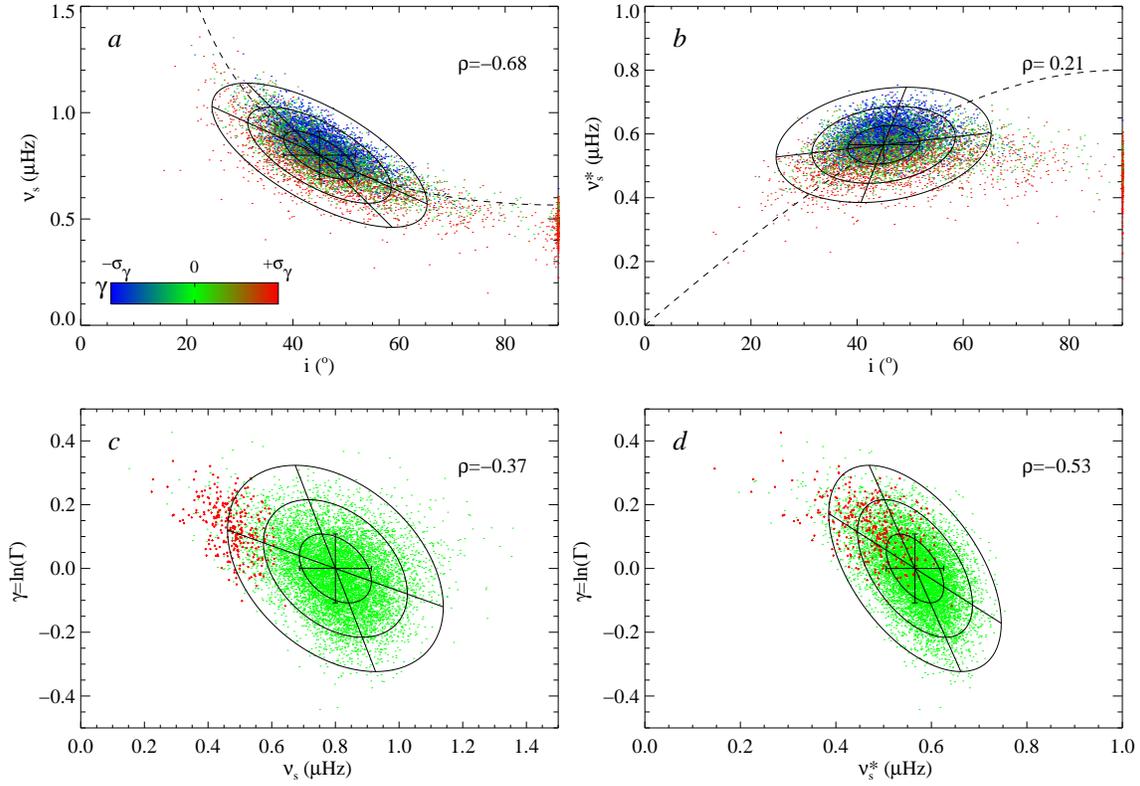}
\caption{\textit{Top}: correlations between the angle $i$ and the splitting $\nu_s$ (\textit{a}) or the projected splitting $\nu_s^{*}$ (\textit{b}). Dashed line indicates $\nu_s^{*}={}$\modif{const} (\textit{a}) or $\nu_s={}$\modif{const} (\textit{b}). Point color indicates the fitted value of linewidth $\gamma$. The color scale is saturated at $\pm\sigma_\gamma$ ($\sigma_\gamma=0.108$).
\textit{Bottom}: correlations between the linewidth $\gamma$ and the splitting $\nu_s$ (\textit{c}) or the projected splitting $\nu_s^{*}$ (\textit{d}). Red \modif{dots} indicate fits where $\tilde i > 89\degr$.
For each plot, $1\sigma$ error bars, \modif{ellipses of errors} (with $k=1$, 2 and 3; see text) and regression lines are deduced from \modifb{the} mean Hessian, assuming normal distribution\modifb{s} of results. \modif{The value of the correlation coefficient $\rho$ deduced from the mean Hessian is also shown}.
}\label{Fig:cor}
\end{figure*}

Let us assume as an example for the remaining of this paper a star characterized by $i=45\degr$, $\nu_s=0.8\unit{\mu Hz}$ ($2\Omega_{\sun}$), $\Gamma=1\unit{\mu Hz}$, and $\beta=1/20$ for $l=1$ modes and $1/10$ for $l=2$. \modif{This corresponds to signal-to-noise ratios about five times smaller than those observed for the Sun with VIRGO.} We consider here a single $l=1$ multiplet observed \modifb{for} 4 years -- that corresponds to Kepler long observation runs -- and perform a MC simulation with 10\,000 realizations. The theoretical errors and correlations have been deduced from the theoretical Hessian. Figure~\ref{Fig:cor} shows both MC and theoretical results and allow us to verify the consistency of both computations for several pairs of parameters. MC results are shown as \modifb{clouds of points}. From the theoretical Hessian, we have derived the error bars, but also the \modif{ellipses of errors} and the regression lines to make visible \modifb{the} correlation.
We \modifb{recall} that an ellipse of errors is, for a 2D normal distribution, an isoprobability curve given by the equation $Q(\lambda_i,\lambda_j)=k^2$ where $Q$ is the quadratic
\begin{eqnarray}
\lefteqn{Q(\lambda_i,\lambda_j)=}&&\nonumber\\
&&\frac{1}{1-\rho_{ij}^2}
\left[
\frac{(\lambda_i-\mu_i)^2}{\sigma_i^2}
-\frac{2\rho_{ij}(\lambda_i-\mu_i)(\lambda_j-\mu_j)}{\sigma_i\sigma_j}
+\frac{(\lambda_j-\mu_j)^2}{\sigma_j^2}
\right]
\end{eqnarray}
where $\mu_{i}$, $\sigma_{i}$ are the mean values and the standard deviations and $\rho_{ij}$ the correlation coefficient (see \S\ref{SSec:MLE}). \modif{The number $k$ is a real constant -- generally integer -- and defines the confidence level: the probability to get a point inside the ellipse is \mbox{$1-\exp(-\frac{1}{2}k^2)$}.}

As for the Sun, the pair of parameters $(A,\Gamma)$ are strongly correlated such $\tilde A\tilde\Gamma\approx{}$\modif{const} (not plotted).
We have found  a correlation coefficient $\rho_{a\gamma}=-0.91$ for our example. This coefficient is almost insensitive to the values of the inclination and the splitting. By exploring different values of $i$ and $\nu_s$, we always find a correlation with an absolute value larger than 0.8.
This is, of course, independent of choosing either $\nu_s$ or $\nu_s^{*}$ as a free parameter.

\begin{figure}[!ht]
\centering
\includegraphics[width=7.5cm]{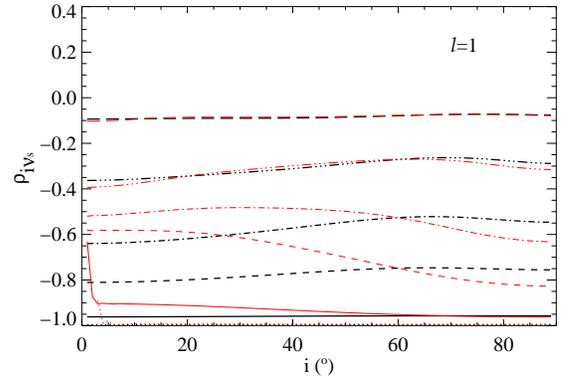}
\caption{Correlation coefficient between $i$ and $\nu_s$ for \modifb{an} $l=1$ mode as a function of $i$, for different values of $x_s$ (see caption of Fig.~\ref{Fig:errnu}). Black lines show errors obtained with the simplified Hessian, thin red lines are obtained by inverting the full Hessian.}
\label{Fig:corispl}
\end{figure}

The second strong correlation we observe is between $i$ and $\nu_s$ (Fig.~\ref{Fig:cor}a).
We recover the result of \citet{BallotG06}: for low rotation rate, $(i,\nu_s)$ are correlated such that $\tilde\nu_s\sin\tilde i\approx{}$\modif{const}.
In our example, the correlation coefficient $\rho_{i\nu_s}=-0.68$, and $|\rho_{i\nu_s}|$ is even greater than 0.9 for $\Omega=\Omega_\odot$. By using $\nu_s^{*}$ instead of $\nu_s$ (Fig.~\ref{Fig:cor}b), the correlation decreases: $\rho_{i\nu_s^{*}}=0.21$ for our example, and we have obtained $\rho_{i\nu_s^{*}}\approx-0.1$ for $\Omega=\Omega_\odot$.
It is important to recall that it is true at low rotation rate only;
when $x_s \gg 1$, the situation is opposite, $i$ is more correlated with $\nu_s^{*}$ than with $\nu_s$.
Figure~\ref{Fig:corispl} illustrates, for a $l=1$ mode, the change of $\rho_{i\nu_s}$ from -1 to 0 \modifb{as $\nu_s$ increases}. We get similar results for $l=2$.
The correlation coefficients are deduced from both the full theoretical Hessian and the simplified one proposed in \S\ref{SSec:Errnui}. We \modifb{note} the limitation of the latter calculation for intermediate values of $\nu_s$: for these configuration, that gives the good order of magnitude for $\rho_{i\nu_s}$, but an incorrect dependency on $i$.

Figures~\ref{Fig:cor}c and \ref{Fig:cor}d illustrate the correlation of $\Gamma$ with $\nu_s$ and $\nu_s^{*}$. We can also see it by looking at the color gradient, which appears on the upper panel. This correlation is significant and is larger with $\nu_s^{*}$ \modifb{than with $\nu_s$}.
That is understandable: the fitting technique has problems \modifb{distinguishing} between the broadening due to splitting and \modifb{that} due to the natural linewidth. We have verified that the correlation \modifb{decreases strongly} when $x_s \gg 1$.

Let us finally notice the dense group of points with $\tilde i \approx 90\degr$ obtained with the MC simulation (Figs.~\ref{Fig:cor}a and b). The phenomenon is discussed in \S~\ref{Sec:Lock}. These peculiar fits are indicated by \modif{red dots} on Figs.~\ref{Fig:cor}c and \ref{Fig:cor}d and we clearly note that, due to correlations, this set of points is shifted -- this indicates a noticeable bias especially on $\nu_s$. We \modifb{notice} also these points are organized along the regression line at constant $\nu_s$, \modifb{which} indicates that, when $i$ is locked at 90\degr, $\nu_s$ is almost blocked around \modifb{an} underestimated values; that is another consequence of the ($i$,$\nu_s$) correlation.


\citet{BallotG06} suggest \modifb{using} $\nu_s^{*}$, instead of $\nu_s$, at low rotation rate, to improve some averages and avoid some effects of correlations: for instance to estimate a mean splitting, $\langle \tilde\nu_s^{*}\rangle/\langle \sin \tilde i\rangle$ can be a better
 estimator than $\langle \tilde\nu_s\rangle$. We have verified \modifb{this} point with
a MC simulation \modifb{of} our example. We have performed 2000
realizations of 10 modes (5 $l=1$ modes and  5 $l=2$ modes).
The distribution of $\langle \tilde\nu_s^{*}\rangle/\langle \sin \tilde i\rangle$ compared $\langle \tilde\nu_s\rangle$ \modifb{exhibits} a smaller dispersion (0.10 against 0.15\unit{\mu Hz} in our case) with a reduced tail at high $\nu_s$.

One must also take into account the existing correlations when one computes derived variables \modifb{such as} the total power of a mode $P$. We denote $p=\ln P$.
A naive estimation of $\sigma_p^2$ is $\sigma_a^2+\sigma_\gamma^2\approx 0.47^2$ (for our example with $T={}$6 months). However due to the correlation, we have to consider the covariance to recover a correct error:
$\sigma_p^2=\sigma_a^2+\sigma_\gamma^2+2\sigma_{a\gamma}\approx 0.15^2$ (this value is in agreement with those found with a MC simulation, see \S\ref{Sec:Biases} Fig.~\ref{Fig:bias}). This point seems obvious but is frequently forgotten.

\modifb{Lastly}, when it is possible to determine the angle $i$ with other techniques, one can fix it and thus one stands in a position similar to the helioseismology, which is more confortable.
However, we have to keep in mind that, due to the complex correlations between all of the parameters, an angle $i$ \modifb{assigned} to a wrong value introduces biases in the determination of other parameters and so introduces systematics which should be estimated on a \modifb{case by case} basis.

\section{Biases for $\vec{\lambda}$ and $\vec{\sigma}$}\label{Sec:Biases}

\begin{figure}[!ht]
\centering
$T=6$ months\hspace*{3cm}$T=4$ years\\
\includegraphics[width=4.5cm]{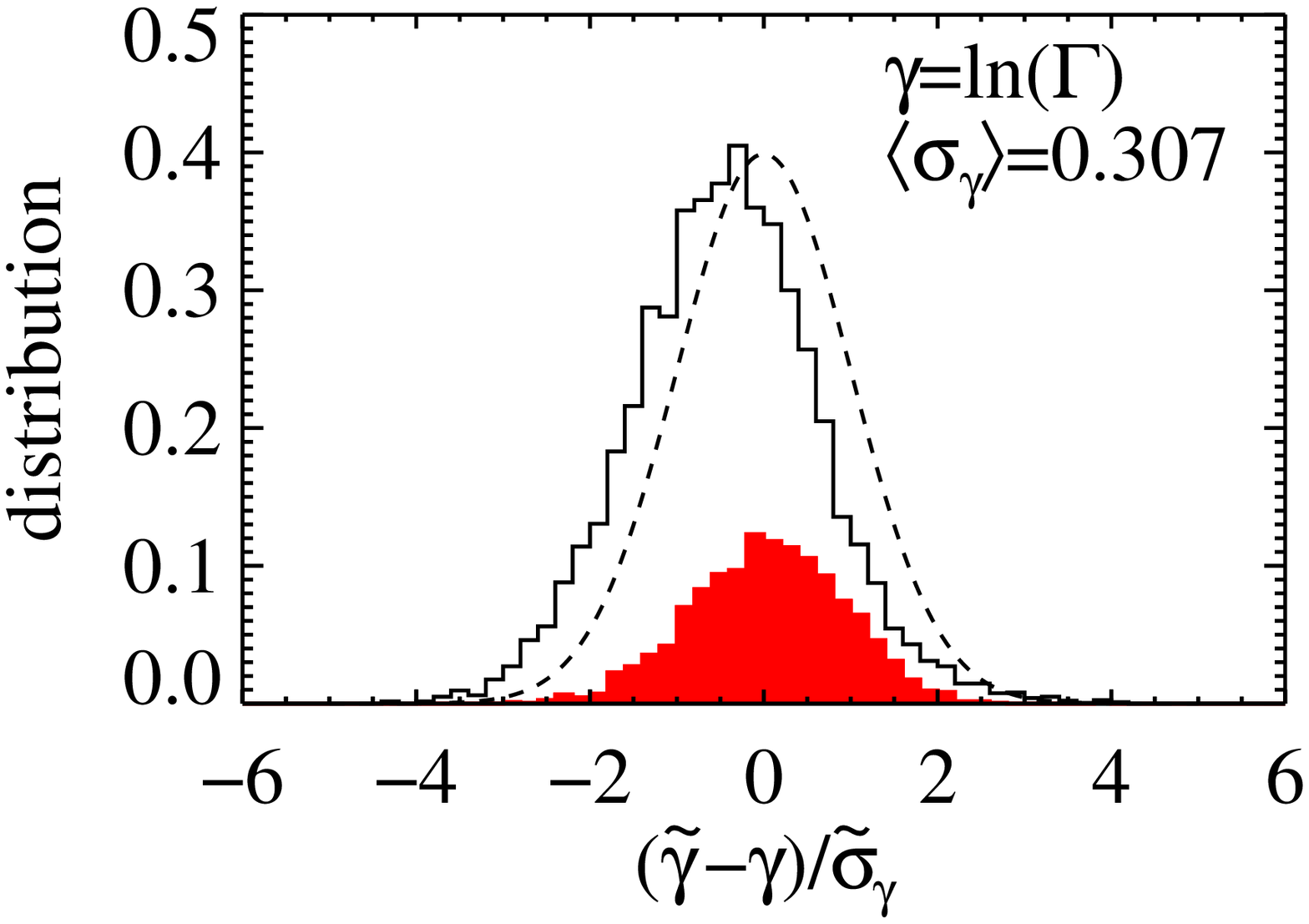}\hspace*{-5mm}%
\includegraphics[width=4.5cm]{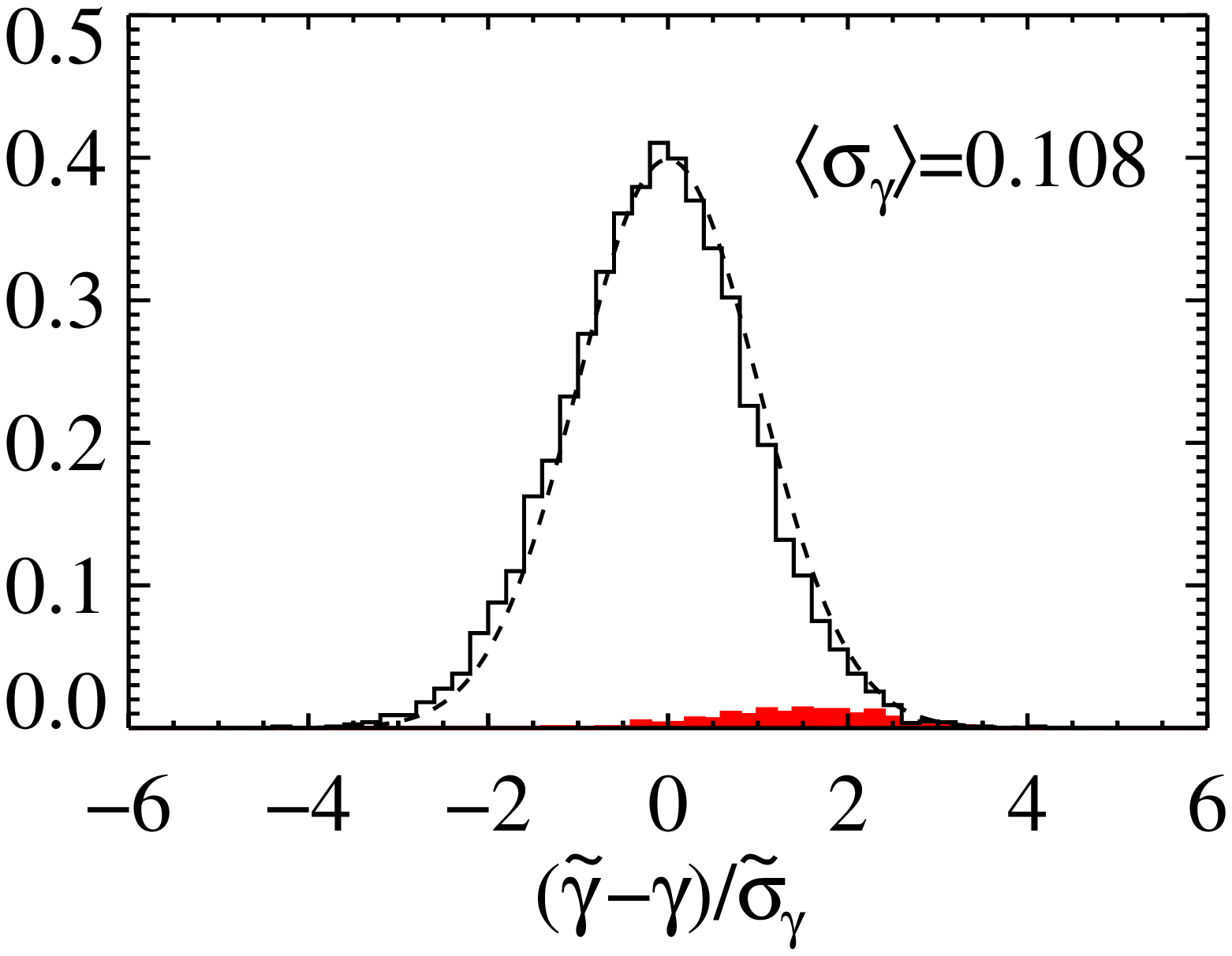}\\
\includegraphics[width=4.5cm]{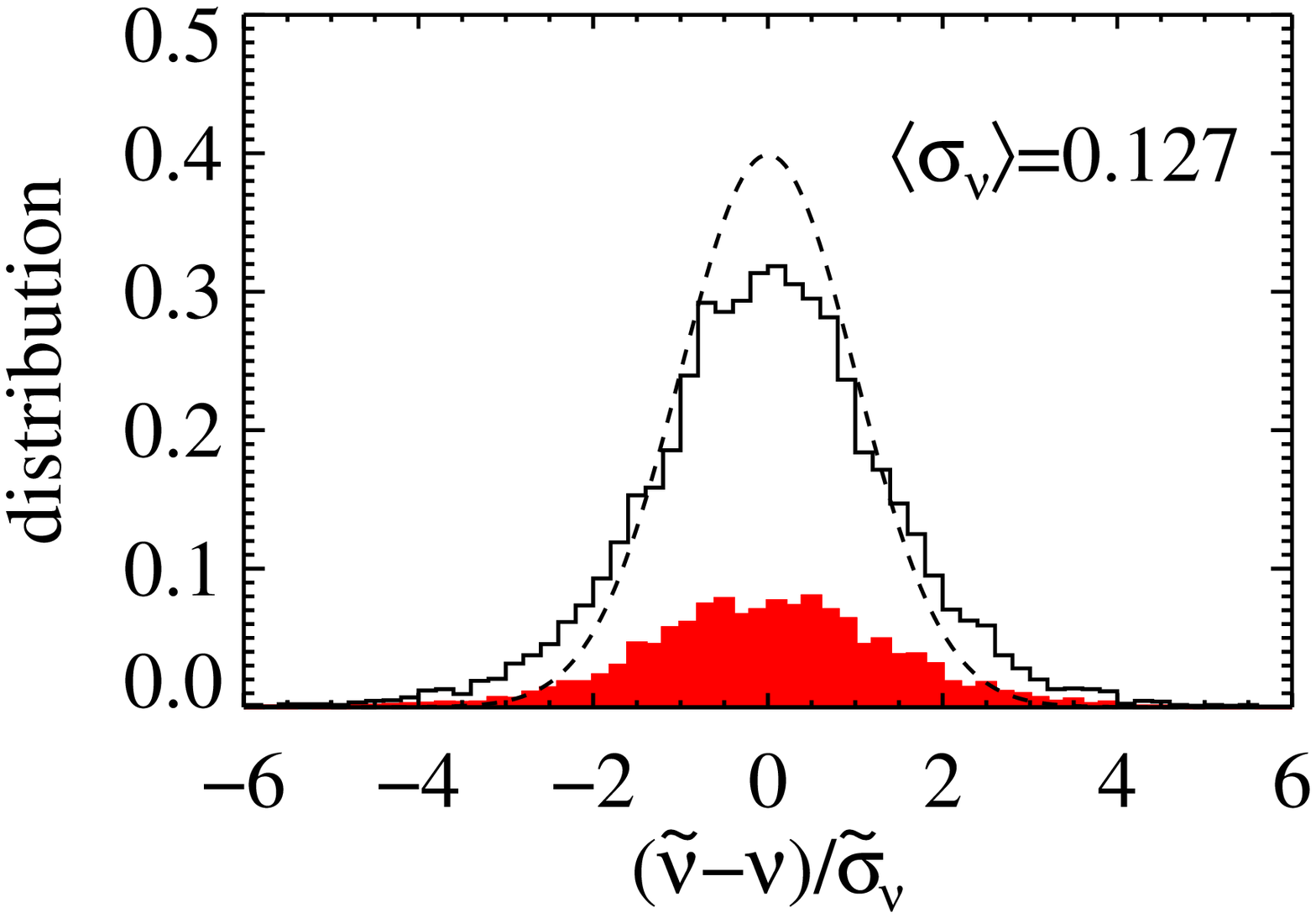}\hspace*{-5mm}%
\includegraphics[width=4.5cm]{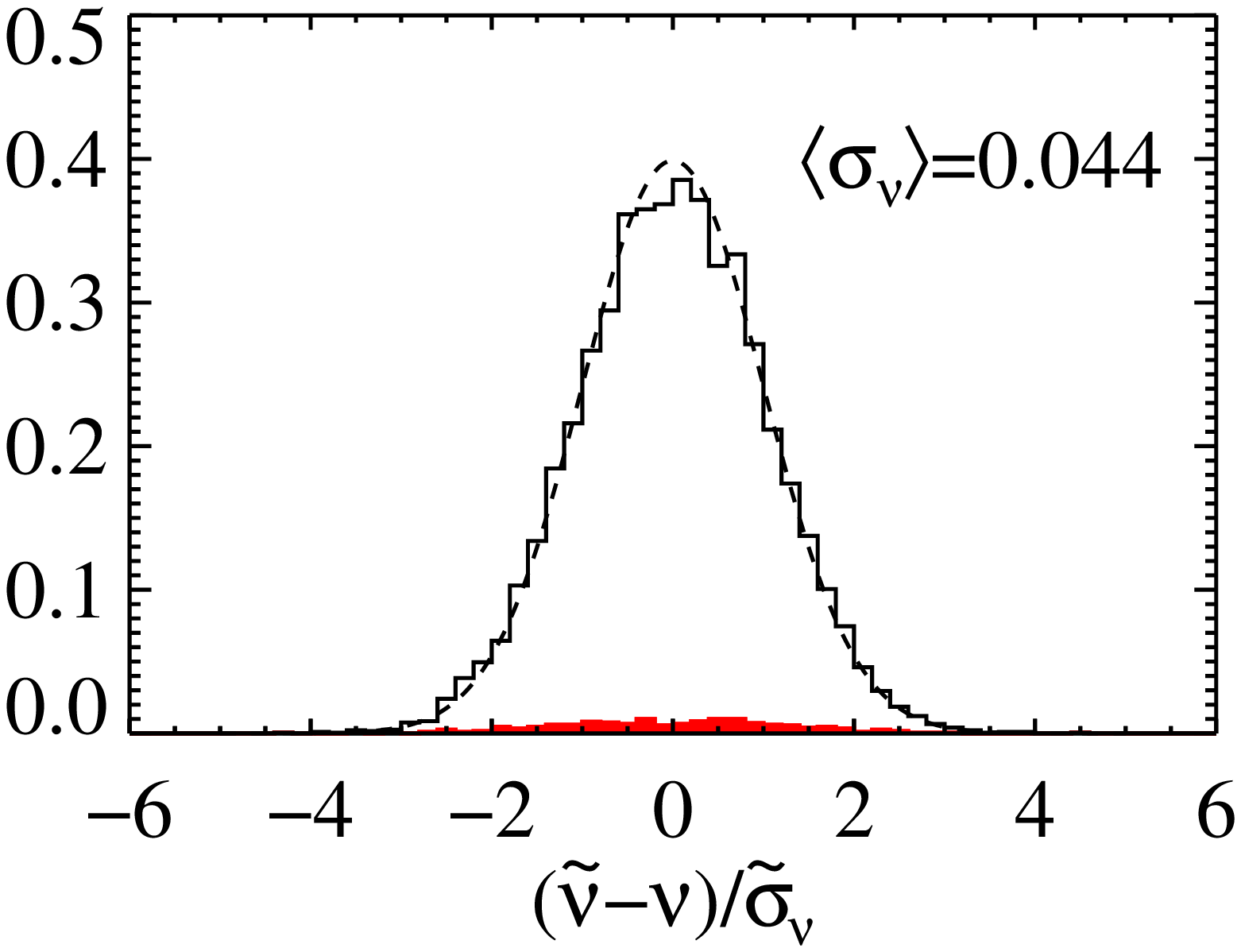}\\
\includegraphics[width=4.5cm]{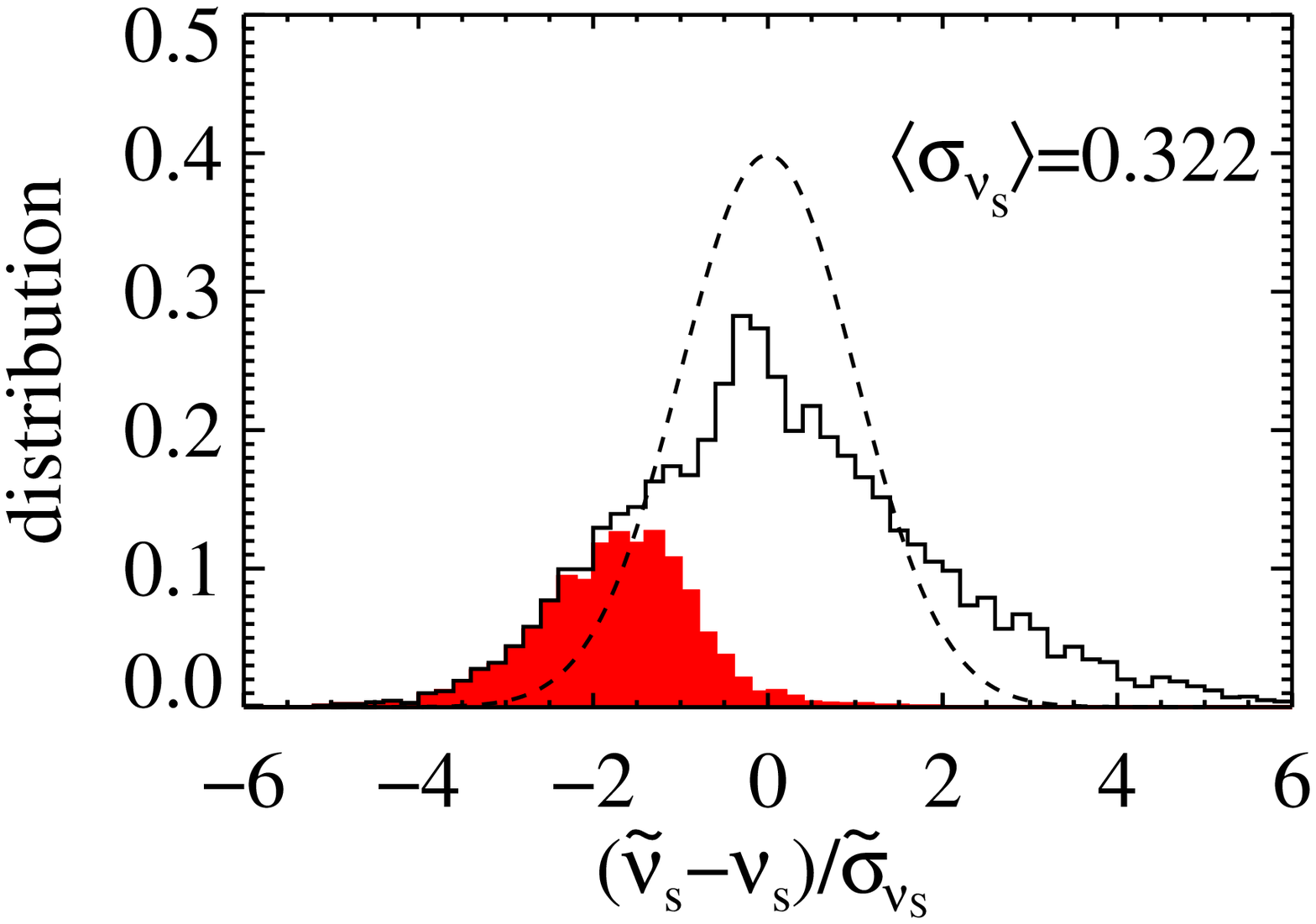}\hspace*{-5mm}%
\includegraphics[width=4.5cm]{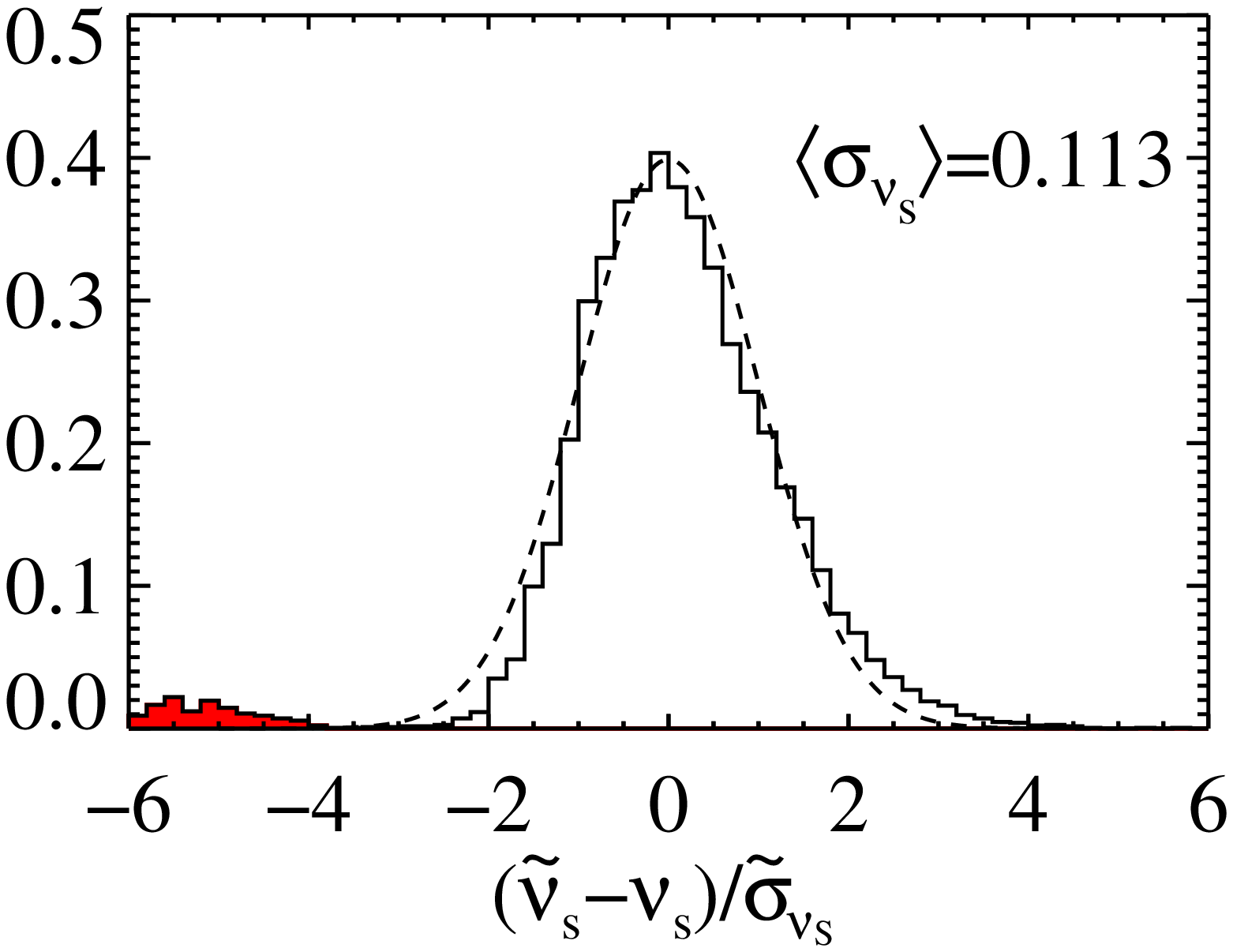}\\
\includegraphics[width=4.5cm]{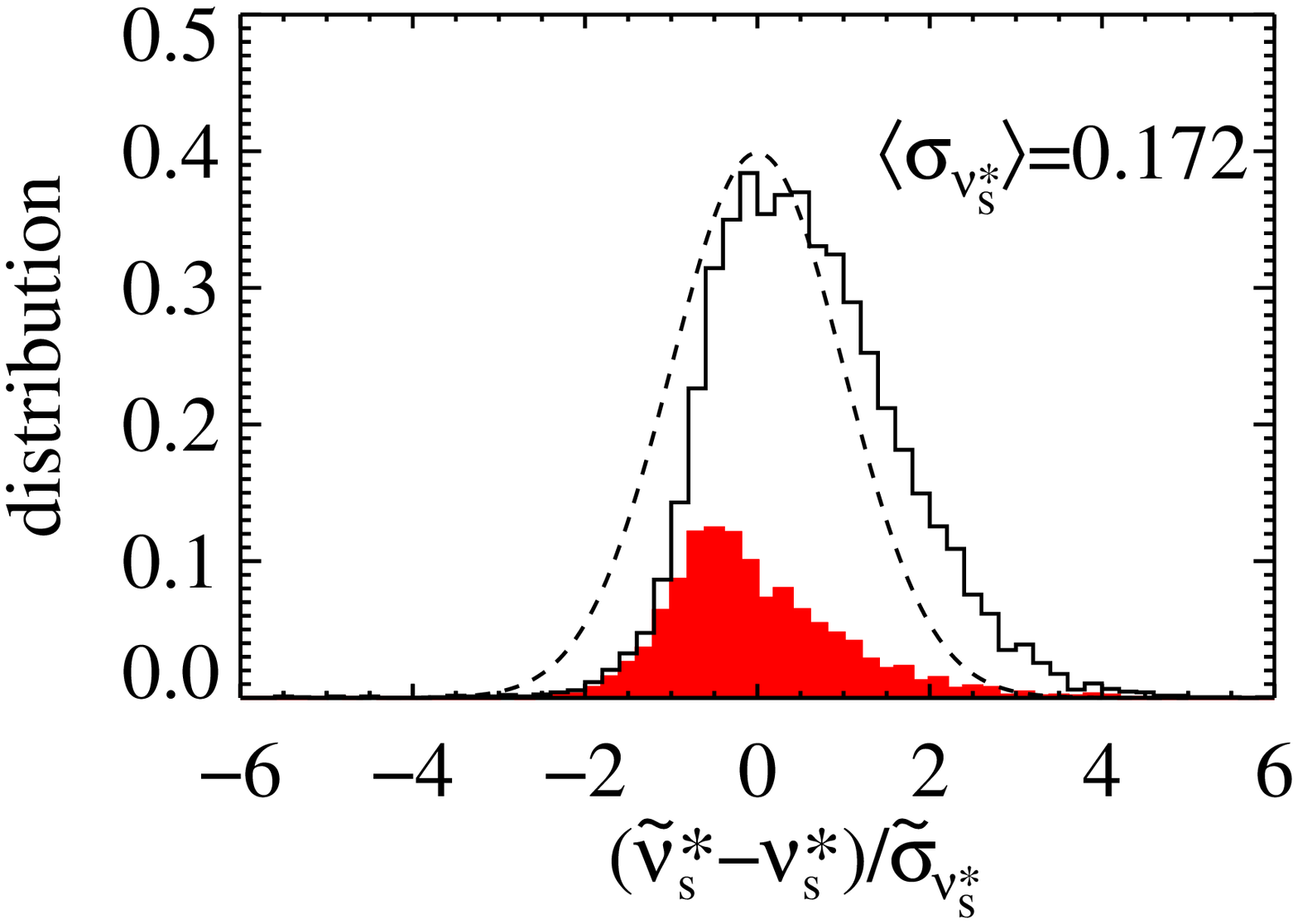}\hspace*{-5mm}%
\includegraphics[width=4.5cm]{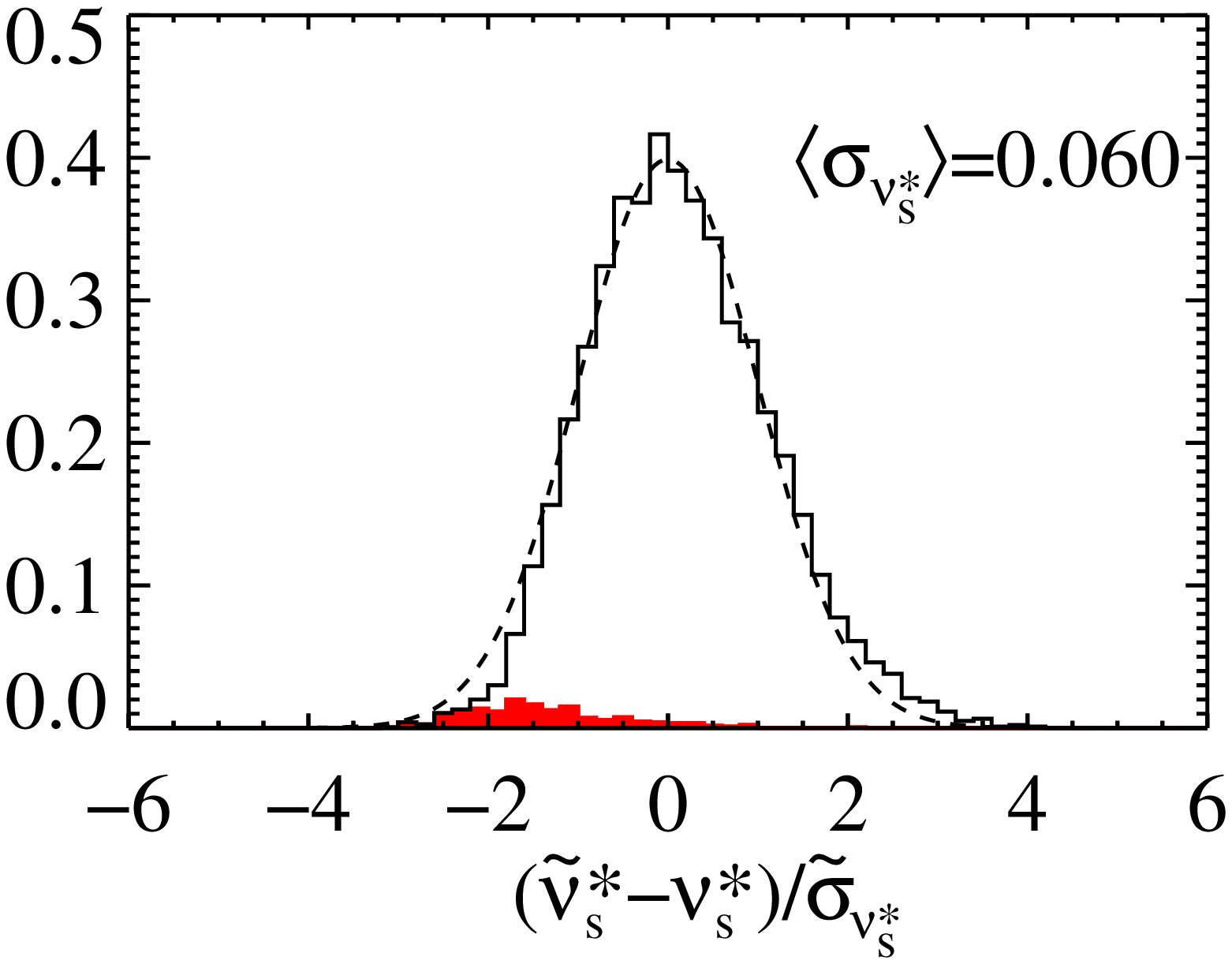}\\
\includegraphics[width=4.5cm]{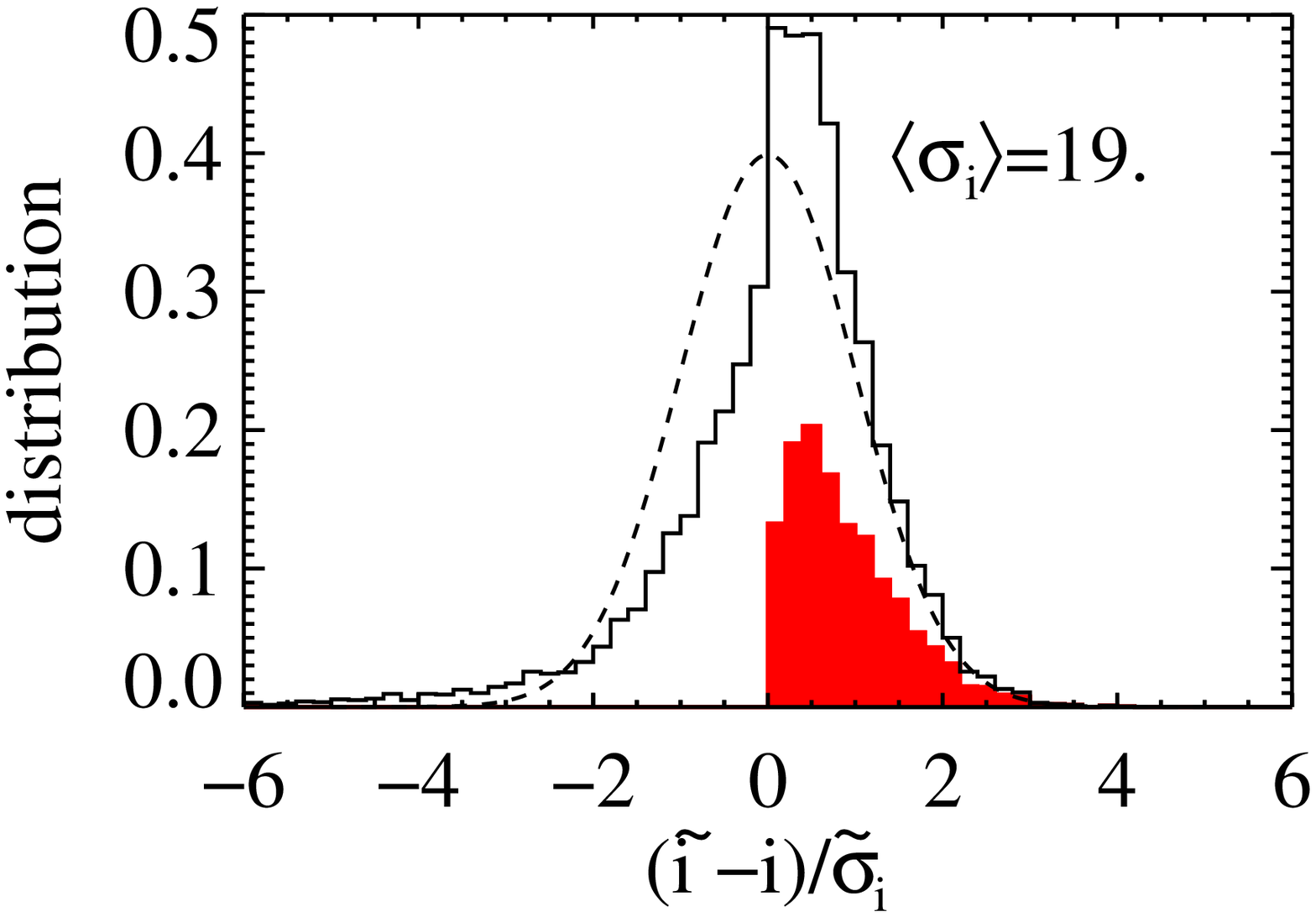}\hspace*{-5mm}%
\includegraphics[width=4.5cm]{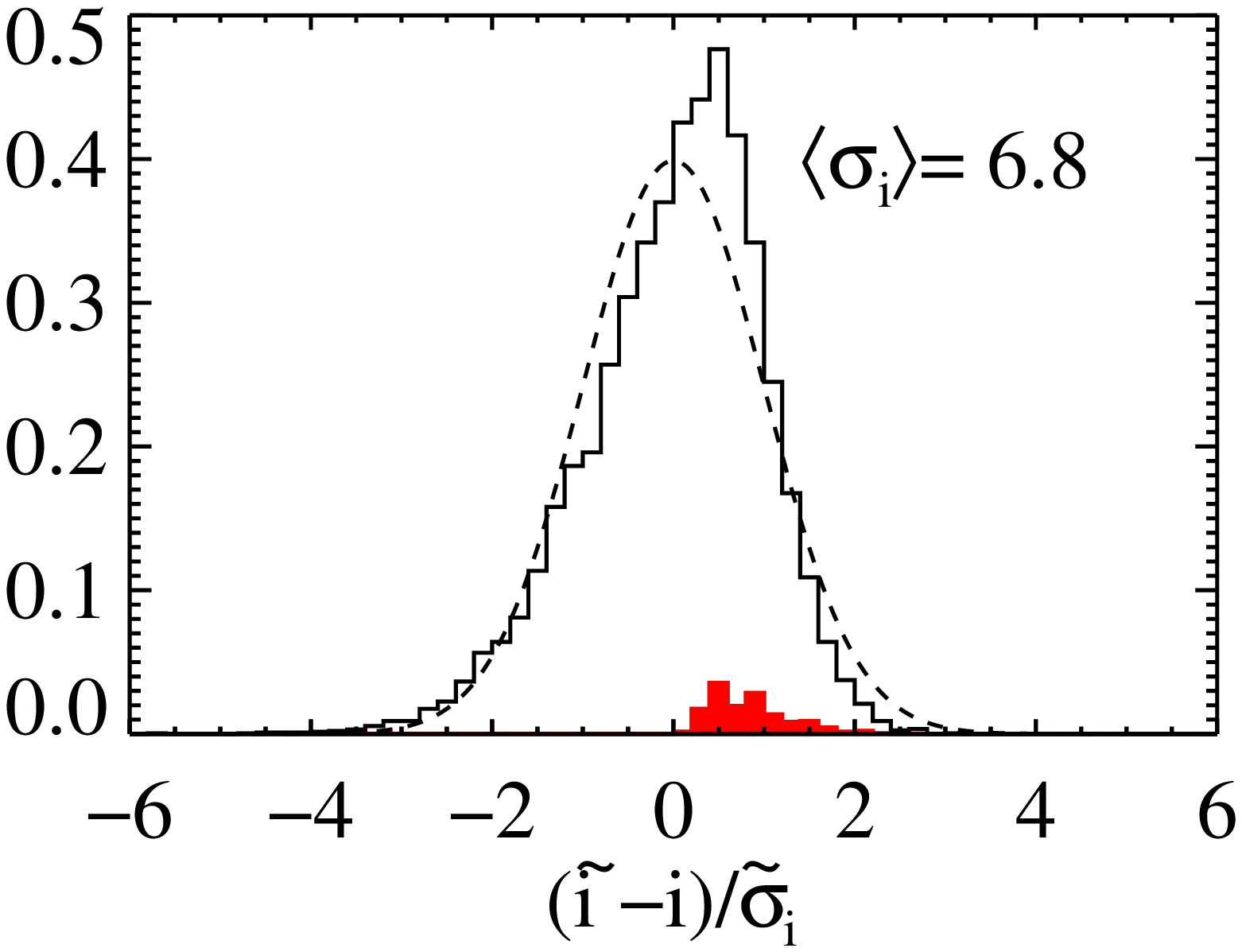}\\
\includegraphics[width=4.5cm]{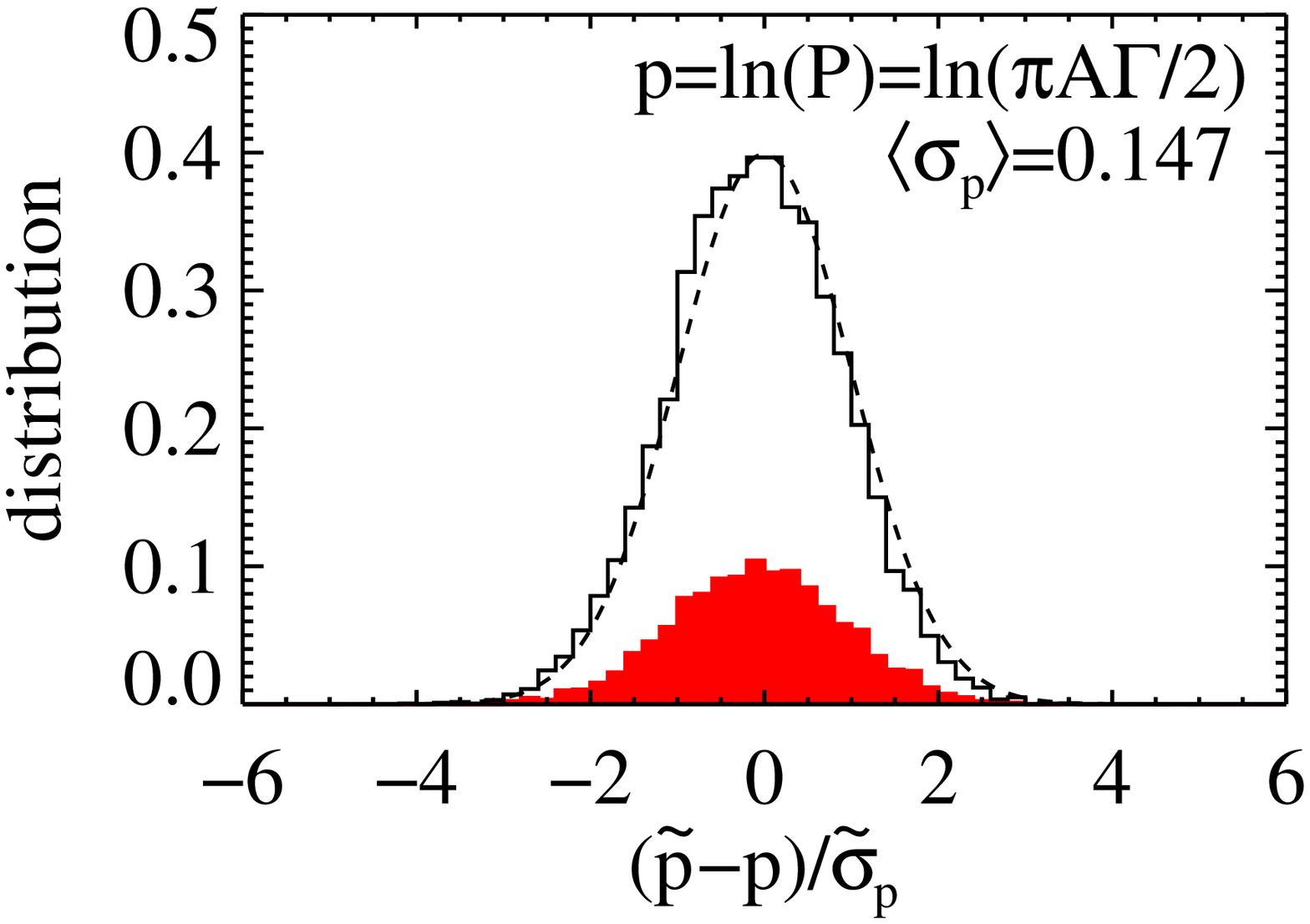}\hspace*{-5mm}%
\includegraphics[width=4.5cm]{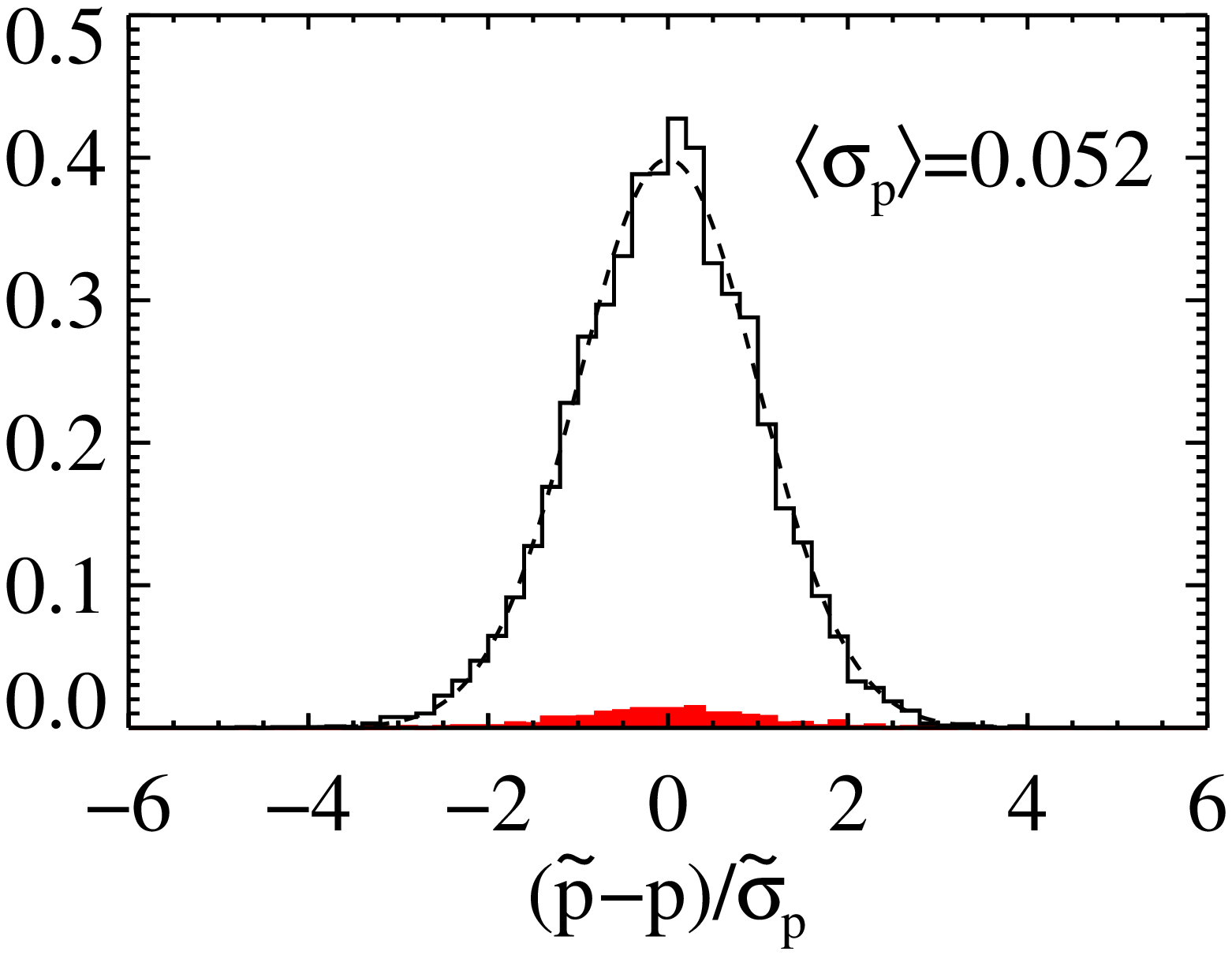}
\caption{Distribution of real errors, $\tilde\lambda_j-\lambda_j$, normalized by the estimated errors $\tilde\sigma_j$ for different parameters (from top to bottom: linewidth, central frequency, splitting, projected splitting, angle and mode power), for two different durations: $T=6$ months (\textit{left}) and 4 years (\textit{right}). On each plot, the filled histogram is the distribution when only realizations with $\tilde i > 89\degr$ are taken into account, dash\modifb{ed} line is the standard normal distribution, and
the mean value $\langle\tilde\sigma_j\rangle$ is the error deduced from the theoretical Hessian, quoted in $\mathrm{\mu Hz}$ for $\nu_0$, $\nu_s$ and $\nu_s^{*}$, and in degree\modifb{s} for $i$.}\label{Fig:bias}
\end{figure}

We know that asymptotically, for long observation times, our estimators of parameters $\tilde\lambda_j$ and their associated errors $\tilde\sigma_j$ are non-biased \citep[e.g.][]{AppourchauxG98}.
However, because observation durations are finite, biases can appear both on $\tilde\lambda_j$ and $\tilde\sigma_j$.
To study this point, we have used the results of the MC simulation described in the previous section. Asymptotically, for $T\rightarrow \infty$, due to the Central Limit Theorem (CLT), fitted parameters follow normal laws; thus $(\tilde\lambda_j-\lambda_j)/\tilde\sigma_j$ asymptotically follows a standard normal law.
Figure~\ref{Fig:bias} shows, for our simulated case, the distribution of the ratio between real errors $\tilde\lambda_j-\lambda_j$ and estimated error bars $\tilde\sigma_j$ for different parameters and for two different observation durations: 6 months (close to CoRoT long runs) and 4 years (Kepler long runs). These plots allow us to see potential biases on parameter determinations or over/underestimations of errors, by comparing the position and the shape of the distribution relative to the standard normal law.

First, the \modif{logarithmic linewidth} $\gamma$ is slightly biased after 6 months of observations: fitting tends to underestimate its value. However, the distribution spread is close to the expected one, that indicate\modifb{s} correct estimations of errors $\tilde\sigma_\gamma$.
The distribution of \modif{logarithmic heights} $a$ (not plotted) is very similar to those of $\gamma$, \modifb{and in particular} we find a similar small bias for short time\modifb{s}, but in the opposite \modifb{sense}.
As a consequence of these opposite biases, the sum of both, i.e. the mode power $p$, is almost non\modifb{-}biased, even for 6-month observations and its distribution is symmetric and almost normal.
If we turn now to the central frequency, $\nu_0$, we notice no bias for the fitted values $\tilde\nu_0$, but for 6-month observations, the errors are generally underestimated, as shown by the \modifb{extended} wings of the distribution.

Six months are generally too short to verify the CLT, especially for splitting -- projected or not -- and angle. The distribution of $\nu_s$ is non-gaussian and is moreover highly spread, that indicates $\tilde\sigma_{\nu_s}$ are \modifb{significantly} underestimated. The distributions of $i$ and $\nu_s^{*}$ are less spread than the previous one, but exhibit clear asymmetries, due to biases for both estimated parameters and errors.

After 4 years, all of the distributions are close to normal. Nevertheless, the angle distribution is still asymmetrical and exhibits a slight bias. The distributions of splitting and projected splitting \modifb{also show} a very small asymmetry but become very close to the standard gaussian, that is a noticeable change mostly for $\nu_s$.

Let us focus now on the realizations for which fits have been locked at 90\degr. First, we \modifb{note that} the number of such cases is reduced for $T=4$~years compared to 6 months.
Due to correlations, when the fitting converges to $\tilde i={}$90\degr, all of the parameters are expected to be biased; there are few exceptions: the central frequency, because it is not correlated with the other parameters (\S\ref{SSec:Errnu0}), and the mode power, \modifb{a} derived parameter which appears to be almost non\modifb{-}correlated with the angle. Another manner to understand this fact is that, since $i$ only acts on the shape of the multiplet, it does not modify its integral (i.e. the power $P$) or its center of gravity (i.e. $\nu_0$).
All other parameters are affected, and the bias becomes significantly large for $\nu_s$ (greater than $5\tilde\sigma$) when $T={}$4~years, \modifb{which} is visible as a bump on the left of the plot. It has already been mentioned in \S\ref{Sec:Correl} and is visible on Fig.~\ref{Fig:cor}c as a set of crosses shifted \modifb{to} the left toward low splitting by about $3\sigma$. These two results seem in contradiction at first sight, but \modif{$\sigma$ has a different meaning in Fig.~\ref{Fig:cor}c and Fig.~\ref{Fig:bias}. From Fig.~\ref{Fig:cor}c we learn that, when locking occurs, the mean bias for $\nu_s$ is 3 times the \emph{mean} standard deviation since red dots cluster around the outer ellipse of errors corresponding to the 3$\sigma$ level, whereas Fig.~\ref{Fig:bias} shows that, in such situations, the mean bias for $\nu_s$ is 5 times the error bar \emph{estimated} for the given case}. This indicates that, \modifb{when locking occurs}, the error bars for $\nu_s$ also are underestimated.

\section{Fit locking at \mbox{\boldmath$i=90\degr$\unboldmath}}\label{Sec:Lock}

Previously, we have seen that a non\modifb{-}negligible proportion of fittings converge toward a solution with $\tilde i = 90\degr$. This phenomenon has been first observed by \citet{GizonS03}. We summarize in Table~\ref{Tab:lockrate} the fraction of such fits for different situations.
Once again, we have considered here our \modif{sample star}, for several observation durations from 3 months to 4 years. We have fitted single $l=1$ modes, pairs of modes $l=0$ and 2, and also sequences of three modes $l=0$, 2 and 1. $l=1$ modes have been fitted using ($i,\nu_s$) or ($i,\nu_s^*$) as free parameters, and, in one case, all \modifb{of} the other parameters have been fixed to their exact values.

\begin{table}
\centering
\caption{Fraction of fitting locked at $i=90\degr$ for different observation times and different sequences of fitted modes}\label{Tab:lockrate}
\begin{tabular}{lccccc}
\hline\hline	
$T$ & 3m  &  6m & 1y & 2y & 4y\\
\hline
$l=$1 ($i,\nu_s$)      & 29\% & 26\% & 18\% & 10\% &  3\%  \\
$l=$1 ($i,\nu_s^{*}$)  &      & 24\% &      &      &  3\%  \\
$l=$1 ($i,\nu_s$)$\dag$&      & 22\% &      &      &       \\
$l=$0\&2 ($i,\nu_s$)   &      & 18\% &      &      & 0.2\% \\
$l=$0,2\&1 ($i,\nu_s$) &      &  8\% &      &      & 0.2\% \\
\hline\hline
\multicolumn{6}{l}{$\dag${\scriptsize($B,A,\Gamma,\nu_0$ fixed)}}
\end{tabular}
\end{table}

We can draw several conclusions.
As we expect, the fraction decreases when $T$ increases; thus, for Kepler-like observations (4~years), there is almost no problem. Second, the results are almost the same by fitting either ($i,\nu_s$) or ($i,\nu_s^*$), there is no improvement by using $\nu_s^*$. We notice also that even when we impose the input $A$, $\Gamma$ and $\nu_0$,  the part of bad fittings is not significatively reduced. This indicates the problem is intrinsic and not just a convergence failure as explain\modifb{ed} below.
Next, since they have more components, $l=2$ modes give slightly better results, though their S/N ratio is lower. However, for $l=2$, another problem appears: in numerous configurations, the fitting hardly distinguishes between a low $\nu_s$ with a high $i$, or a lower $i$ with a  $\nu_s$ \modifb{two times larger}, due to a confusion between $m=\pm 1$ and $\pm 2$ components \citep[See also][]{GizonS03}. Fitting simultaneously $l=1$ and 2 modes allows us to solve this ambiguity. However, although the best results are obtained with sequences of three modes, there are still almost one bad fitting out of ten, for 6-month observations (CoRoT-like observations) -- that is not negligible.

\begin{figure}[!ht]
\centering
\includegraphics[width=7.5cm]{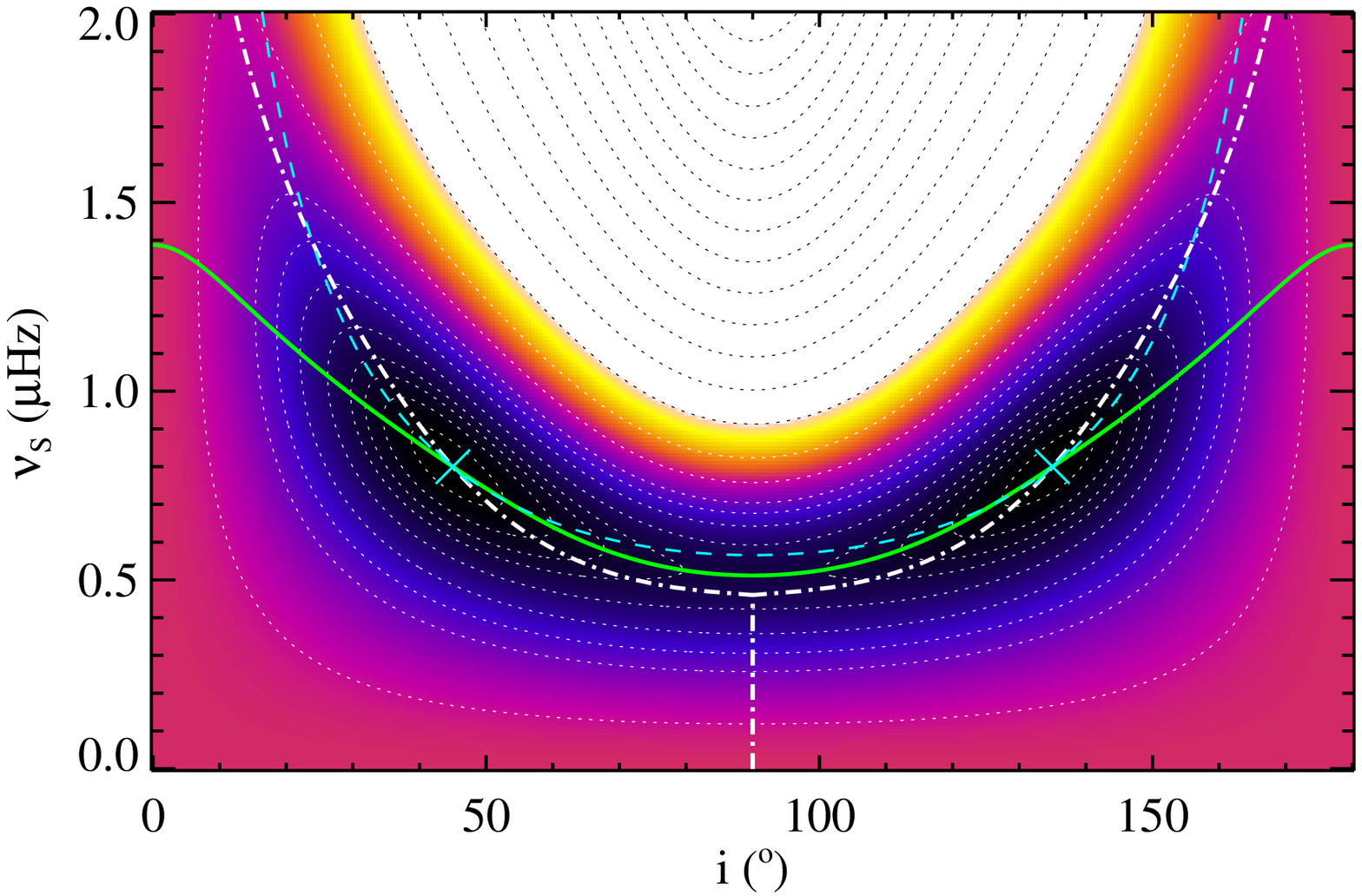}\\
\includegraphics[width=7.5cm]{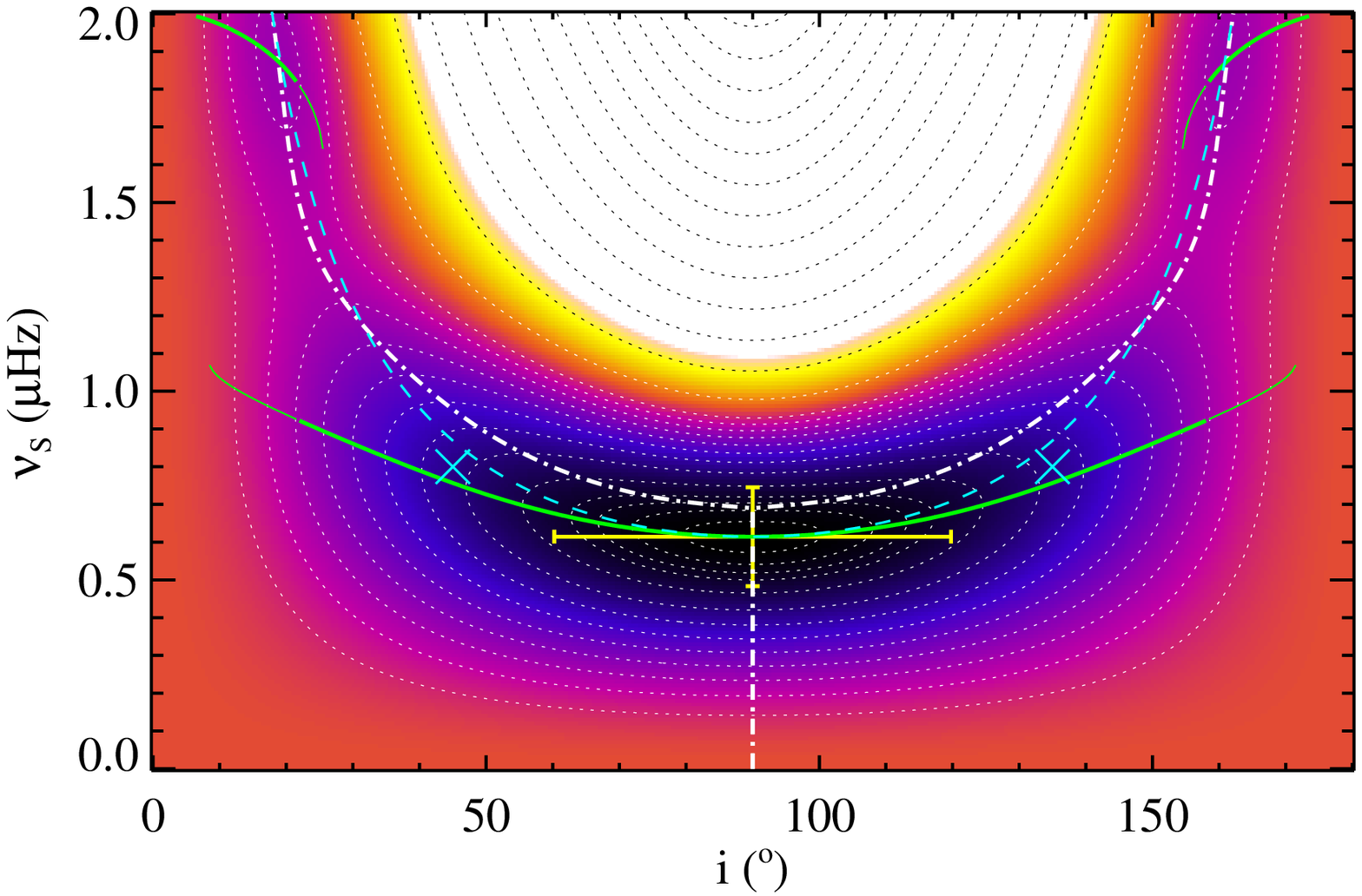}
\caption{Maps of likelihood $\ell=-\ln {\cal L}$ in the plane ($i,\nu_s$) for a $l=1$ mode. \textit{Top}: map of the mean likelihood. \textit{Bottom}: an example of realization where the locking phenomeoccurs. Black corresponds to the minimum of $\ell$, white to higher values. \modif{Fine white or black dotted lines are contours of constant likelihood}. 
Crosses indicate the input values of $i$ and $\nu_s$ in the spectum; dashed line indicates $\nu_s^*={}$\modif{const}.
Solid lines indicate the minima of $\ell$ along $i$ at fixed $\nu_s$, i.e. points where $\partial_i\ell=0$ and $\partial^2_i\ell>0$; dot-dashed lines indicate the minima of $\ell$ at fixed $i$; thick lines mean that this is a global minimum, fine lines a local one. On \modifb{the} bottom plot, the \modifb{cross} with errors bars indicates the result of the fit.
}\label{Fig:like}
\end{figure}

To understand the origin of the fit locking, we have plotted on Fig.~\ref{Fig:like} (\textit{top}) the mean likelihood in the plane ($i,\nu_s$) for a $l=1$ mode.
The plot is symmetric \modif{with} respect to $\pi/2$ (and $\pi$-periodic).
As expected, minima correspond to the values of ($i,\nu_s$) used to model the spectrum.
Since \modifb{the likelihood}, $\ell$, is even \modifb{in} $i-\pi/2$, a valley along $i=\pi/2$ appears at low splitting, before a fork \modifb{appears} at higher splittings.
Obviously the (black) region around the minimum extends up to the limit of 90\degr, which means that it is quite likely that for \modifb{a} noise realization 
the minimum of the likelihood will end up close to that limit.
It may even end up beyond that limit and in this case because 
of the periodicity of the inclination there is obviously no inclination which could allow the fit to reach such a minimum.
Figure~\ref{Fig:like} \textit{bottom} shows an example of such a situation with a global minimum at 90\degr (and a secondary minimum around 20\degr, with interestingly the same value of $\nu_s^{*}$ since it is also on the dash line).
The fit has then been blocked by the hard limit at 90\degr\ and has stopped there producing this effect of locking at 90\degr. 
A similar effect is described in \citet{ChaplinE07} for a problem of fitting solar-cycle frequency shift or in an ongoing work by the same authors
about the zero-lock\modifb{ing} of $l$=1 splittings for large mode linewidths. All \modifb{of} these problems have a \modif{common} origin: a parameter is quite sensitive to the realization noise
and the minimum of the likelihood function because of the realization noise can take a value which cannot be reached in the parameter range (here 0--90\degr). 
In our case to overcome this lock\modifb{ing} it is sufficient to use a parameter $j=\sin^2 i$ instead of $i$. If at the minimum, $j$ is larger than 1 it means 
that the fit using $i$ would have locked at 90\degr. Of course in this case it is not possible to recover a meaningful inclination. 
Obviously the lock\modifb{ing} at 0\degr\ is less likely to happen according to Fig.~\ref{Fig:like} since the black region is confined far from the 0 limit: in other words, the likelihood surface along $i=0\degr$ is a \modifb{ridge} and not a valley as \modifb{it is} along $i=90\degr$.
As observations \modifb{become} longer there are more and more points to describe the mode profile and the region is then more confined around the real minimum 
and further away from the 90\degr\ limit, the lock\modifb{ing} is then less likely to happen unless the
underlying inclination is close to the limit itself.

\section{Conclusion}\label{Sec:Conclusion}
As pointed out \modifb{previously} \citep{GizonS03,BallotG06} the extra parameter $i$ introduces difficulties in fitting stellar spectra, compared to the solar case. The main problem originates in the blending of components \modifb{within} a multiplet which strongly correlates the inclination with the rotational splitting, making \modifb{it} difficult to disentangle them. For slow rotators, the uncertainties of $i$ and $\nu_s$ are huge (\S\ref{SSec:Errnui}),  because of this. The projected splitting $\nu_s^{*}$ \modifb{appears} to be less \modifb{sensitive} to this, thus has more moderate errors.
Beyond problems induced by correlations, other issues \modifb{such as} the locking of $i$ at 90\degr\ often occur (see \S\ref{Sec:Lock}). The correct derivations of mode heights and linewidths are also dependent on a correct recovering of \modifb{the} inclination.

These issues vanish for stars rotating sufficiently \modifb{rapidly} -- or stars with p modes having sufficiently long lifetimes --, if we disregard other issues which appear \modifb{such as} problems in mode identification or blending between different modes.
Fixing $i$ before fitting allows us to get around these difficulties. However, due to the numerous correlations between the different parameters (see \S\ref{Sec:Correl}), the value of $i$ must be carefully chosen to avoid systematics.

Nevertheless, even though the value of $i$ modifies the error expected for the central frequency $\nu_0$ of a mode (\S\ref{SSec:Errnu0}), it does not introduce any biases as \modifb{long} as the multiplet is symmetric enough (\S\ref{Sec:Biases}). Hence, $\nu_0$ should be \modifb{recovered} in any case, as well as certain derived quantities \modifb{such as} the mode power.

\begin{acknowledgements}
The authors want to thank the International Space Science Institute (ISSI) for
\modifb{supporting a} workshop of the asteroFLAG group%
\footnote{http://www.issi.unibe.ch/teams/Astflag/}, where this work was
started. This work \modifb{was} partially supported by the European Helio- and
Asteroseismology Network (HELAS\footnote{http://www.helas-eu.org/}), a major
international collaboration funded by the European Commission's Sixth
Framework Programme. \modif{The authors also wish to thank the referee, A.~F. Lanza, for his relevant comments which helped to improve the paper.}
\end{acknowledgements}

\bibliographystyle{aa}
\bibliography{ballot}

\end{document}